\documentclass{osa-article}
\usepackage{slashbox}
\pdfoutput=1
\journal{osac}

\articletype{Research Article}

\usepackage{lineno}

\begin{document}

\title{Design, simulation and characterization of integrated photonic spectrographs for Astronomy I: Generation-I AWG devices based on canonical layouts}

\author{Andreas Stoll,\authormark{1,*} Kalaga V. Madhav,\authormark{1}  and Martin M. Roth\authormark{1}}

\address{\authormark{1}Leibniz-Institut f\"ur Astrophysik (AIP), An der Sternwarte 16, 14482 Potsdam, Germany}

\email{\authormark{*}astoll@aip.de} 


\begin{abstract}
We present an experimental study on our first generation of custom-developed arrayed waveguide gratings (AWG) on silica platform for spectroscopic applications in near-infrared astronomy. 
We provide a comprehensive description of the design, numerical simulation and characterization of several AWG devices aimed at spectral resolving powers of 15,000 - 60,000 in the astronomical H-band. 
We evaluate the spectral characteristics of the fabricated devices in terms of insertion loss and estimated spectral resolving power and compare the results with numerical simulations. We estimate resolving
powers of up to 18,900 from the output channel 3-dB transmission bandwidth. Based on the first characterization results, we select two candidate AWGs for further processing by removal of the output 
waveguide array and polishing the output facet to optical quality with the goal of integration as the primary diffractive element in a cross-dispersed spectrograph. We further study the imaging properties 
of the processed AWGs with regards to spectral resolution in direct imaging mode, geometry-related defocus aberration, and polarization sensitivity of the spectral image. We identify phase error control, 
birefringence control, and aberration suppression as the three key areas of future research and development in the field of high-resolution AWG-based spectroscopy in astronomy.
\end{abstract}

\section{Introduction}

Compact, low-cost optical instruments play an increasingly important role in scientific disciplines ranging from biomedical research to space exploration. In astronomy, the idea of an integrated photonic 
spectrograph is particularly appealing for space telescopes, where mass and volume are critical parameters for all subsystems of a mission. For example, the NIRSPEC spectrograph onboard the next 
generation space telescope JWST \cite{Ferruit:12}, whose main science cases are (1) cosmology in the epoch of reionization, (2) the assembly of galaxies, (3) the birth of stars and protoplanetary 
systems, and (4) exoplanets and the origin of life, offers a spectral resolving power of up to $R=2700$ in three wavelength bands between $1\,\upmu$m and $5\,\upmu$m. The instrument has a total 
mass of $196$ kg. Near-infrared (NIR) spectroscopy at higher spectral resolution of up to $R=\lambda/\Delta\lambda=50,000-100,000$, as required by science cases like, e.g. exoplanet atmospheres, 
stellar astrophysics, accretion and outflow of young stellar objects, or chemistry of the interstellar medium, would require a cross-dispersion echelle spectrograph that would be hugely more demanding in 
mass and volume, as can be appreciated from a ground-based example such as the CRIRES instrument at the ESO Very Large Telescope \cite{Kaeufl:04}. It is therefore worthwhile to study the achieveable 
performance in terms of spectral resolution, spectral range, and throughput of a NIR integrated photonic spectrograph towards a future replacement of conventional optics in space applications.
This paper is a first in a series where we begin our study with custom-designed AWGs that were derived from canonical layouts known from the literature, in what follows Gen-I. Paper II will address 
Gen-II AWGs with three-stigmatic-point designs, and Paper III is going to discuss echelle gratings, for comparison with Gen-I and Gen-II AWGs. 

\begin{figure}[!ht]
	\centering
	\includegraphics[width=85mm]{ 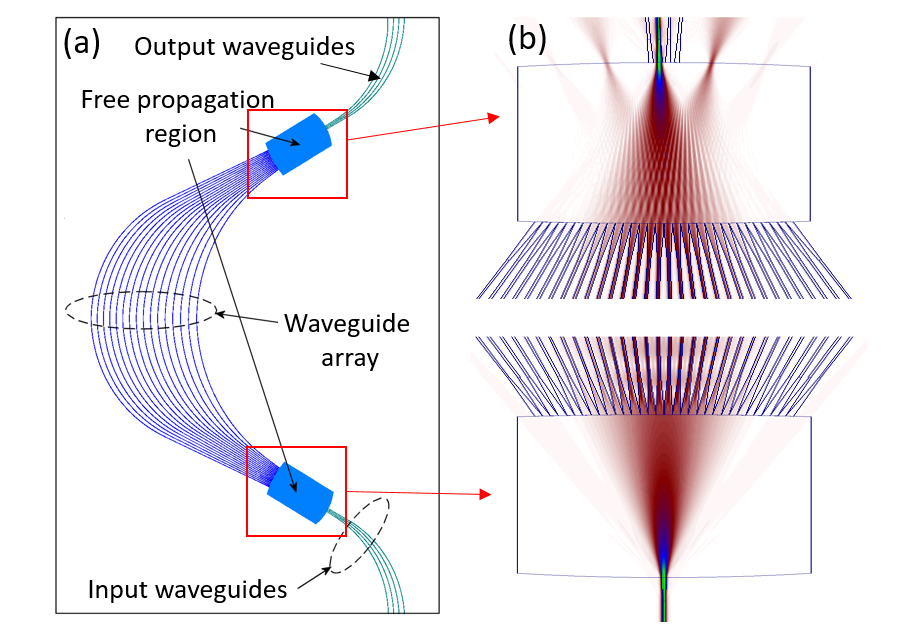}
	\caption{(a) Structure of an AWG device. (b) BPM simulation of light propagation through the input/output star coupler regions.}
	\label{fig:AWG_Intro}
\end{figure}

Arrayed waveguide gratings (AWGs) have already been studied for their properties as primary dispersive elements in integrated photonic spectrographs, especially for 
astronomy \cite{Bland-Hawthorn06, Cvetojevic:09, Cvetojevic:12, Gatkine:17}. Descriptions of the fundamentals of AWGs and their modelling and design are found in the literature 
\cite{Smit88, Takahashi:1990, Dragone:1991, Okamoto:2006, Leijtens:06, Munoz:02}, and the principal structure of a typical AWG device is shown in Figure \ref{fig:AWG_Intro}(a). Figure \ref{fig:AWG_Intro}(b) 
illustrates the propagation of light through the star coupler segments of the AWG, simulated using the beam propagation method (BPM), which was applied in the development of the building blocks for 
the AWG designs presented in this paper.

While commercially available AWGs are well-optimized for dense wavelength division multiplexing (DWDM), they are not suited for use in high-resolution spectroscopy. The practical implementation of high 
resolution AWGs poses significant challenges regarding fabrication tolerances and the resulting impact on the device performance. This work presents an experimental study of custom-designed AWGs on 
a $2\%$ refractive index contrast silica-on-silicon platform, targeting spectral resolving powers $R=15,000...60,000$ in a region of the near-infrared astronomical H-band covering $1500$ nm - $1700$ nm.

The combined requirement of high spectral resolution and large wavelength coverage of a few hundred nm calls for operation in high spectral orders, which inevitably narrows the free spectral range (FSR) of the 
AWG to $10$ nm - $20$ nm. If the spectrum of the input signal covers more than one FSR, the spectral image on the output facet of the AWG will contain multiple overlapping spectral orders with a wavelength
spacing of one FSR between to adjacent orders. A cross-dispersion technique is employed to overcome the FSR limitation by separating the overlapping spectral orders by means of a second dispersive element such 
as a grating or a prism, which is introduced into the optical path after collimation of the AWG output. The collimated beam is dispersed by the second dispersive element in the direction perpendicular to the dispersion 
of the AWG, introducing a slow dispersion axis for the vertical separation of the spectral orders. In an optimized system, the ratio of angular dispersions of the AWG and the cross-disperser should equal the total number
of spectral orders in the operating wavelength range in order to utilize the detector area most effectively. In our instrument design, we are planning to use an Al-coated ruled grating with 600 grooves/mm and a blaze 
wavelength of 1600 nm. 
 \begin{figure}[!ht]
	\centering
	\includegraphics[width=125mm]{ 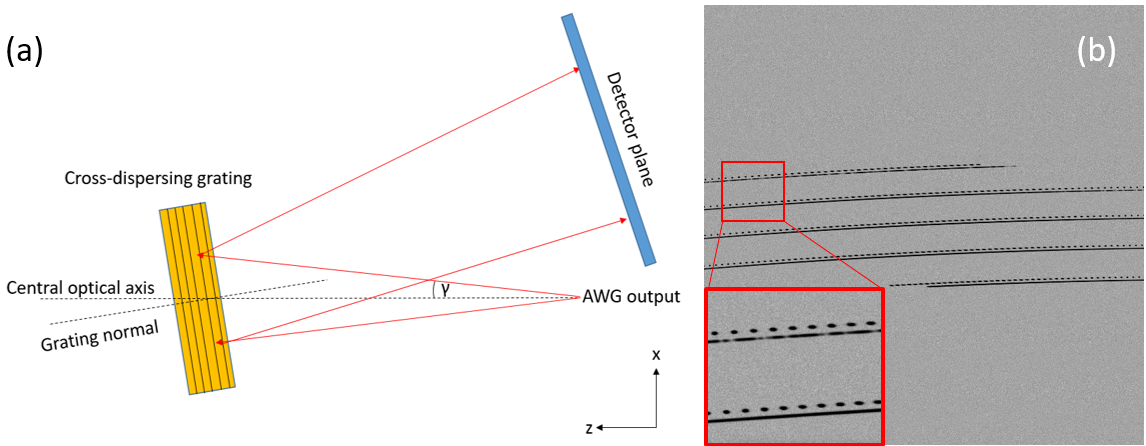}
	\caption{(a) Principal sketch of an AWG spectrograph utilizing cross-dispersion (AWG chip not shown), the detailed opto-mechanical design is presented in \cite{Hernandez:20}. (b) Simulated echellogram as 
imaged by the detector prior to data reduction. Example showing a continuous signal under atmospheric absorption accompanied by a discrete calibration spectrum generated by a frequency comb.}
	\label{fig:crossdisp}
\end{figure}
Figure \ref{fig:crossdisp}(a) shows the principle of a cross-dispersion-assisted AWG spectrograph. The cross dispersion produces a series of spectra called echellogram, shown in Figure \ref{fig:crossdisp}(b), whereby 
each spectrum covers one FSR of the AWG. The example shown in Figure \ref{fig:crossdisp}(b) assumes one single-input AWG device, and a Teledyne H2RG HgCdTe array with $2048\times 2048$ 18 $\upmu$m 
pixels as the detector, leaving most of the detector area unused. A more efficient configuration can be 
achieved by using multiple inputs on a single AWG \cite{Gatkine:18} or by combining multiple single-input AWG devices into a multi-fibre stack \cite{Watson:95} or a possible combination of both. 

The fabrication of functional high-resolution AWGs poses significant challenges in terms of fabrication accuracy. High-resolution AWGs with hundreds of waveguides and optical propagation lengths on the order of cm 
are very susceptible to phase error induced image degradation, leading to fabrication tolerance limited performance \cite{Stoll:20, Gatkine:21}. In earlier publications, we have studied the potential and technical 
limitations of silica-platform AWG, with a focus on the effects of fabrication tolerance related phase errors and the necessity of AWG post-processing by means of UV trimming \cite{Stoll:17, Stoll:20}. In this work,
 we experimentally study several different AWG designs of varying resolving power on a silica-on-silicon platform. We show AWG designs of varying foot-print, study the spectral transmission characteristics of 
fabricated AWG chips, and compare the results to theoretical expectations with the goal to determine the practical feasibility of high-resolution AWGs for spectroscopy.

\section{Design of the basic waveguide structure}
Arrayed waveguide gratings of the Rowland-family are by far the best known type of AWGs due to its origin in the telecommunication industry, which optimized the Rowland AWG design in accordance with its needs. 
In the first iteration of our custom AWG designs, we have implemented the well-known Rowland geometry in various arrangements of waveguide arrays and star couplers, waveguide array sizes between 360 and 722 
waveguides, and theoretical spectral resolutions ranging from $0.1$ nm to $0.026$ nm. The designs were aimed at fabrication on an atmospheric pressure chemical vapour deposition (APCVD) Silica-on-Silicon material platform 
with a core refractive index contrast of $2\%$, which was set by the process used by the foundry. Silica-on-silicon technology was selected as a well-developed and reliable platform in order to minimize the risk of 
failure, as only one fabrication run could be carried out. Furthermore, a low refractive index contrast of $2\%$ allows for very efficient fibre-chip coupling using single-mode high-NA fibres. Certain newer technologies, 
such as silicon nitride TriPleX \cite{Worhoff:15}, are comparable with silica technology in terms of propagation loss. However, due to the large aspect ratio of the SiN core, strong geometrical birefringence is to be expected, 
which is not desirable for our application.

The building blocks of the AWG structure, such as buried channel waveguides, circular waveguide bends, taper waveguides and slab waveguides were independently modelled and simulated using a full-vectorial 3D BPM 
method, implemented in the commercial software RSoft BeamPROP. Simulation of the AWG devices as a whole was carried out using custom-written code based on a scalar diffraction model which relies on the BPM simulation 
results of the individual building blocks to calculate diffraction images at the output end of the AWG star coupler.

The starting point of Arrayed Waveguide Grating modelling is the definition of material parameters of the waveguide structure. The most basic property of an optical waveguide is the material refractive index of the 
waveguide core and surrounding cladding and the refractive index contrast between cladding and core. These quantities are determined by the material platform and fabrication process and ultimately dictate the 
waveguide geometry, namely the single-mode cut-off cross-section of the core, as well as the maximum possible waveguide curvature due to bend waveguide mode leakage. Numerical modelling of basic building blocks 
of an AWG device can be carried out, once the fundamental waveguide structure of the AWG design has been worked out.

\subsection{Design of the waveguide core}
The target material platform of the AWG design is Ge-doped silica-on-silicon with a low refractive index contrast of $2\%$. The waveguide design is implemented as a buried-channel waveguide type consisting of three 
$SiO_2$ layers on top of a Si-substrate. A symmetric aspect ratio of 1:1 is chosen for the waveguide core in order to minimize waveguide mode birefringence. In the first design step, the single-mode cross-section of the 
waveguide core must be determined for the shortest wavelength of the working range of the AWG device by numerical simulation. The single mode condition must be satisfied in the entire wavelength range of operation, 
therefore the single-mode cut-off core cross-section must be determined at $1500$ nm. 
\begin{figure}[!h]
	\centering
	\includegraphics[width=125mm]{ 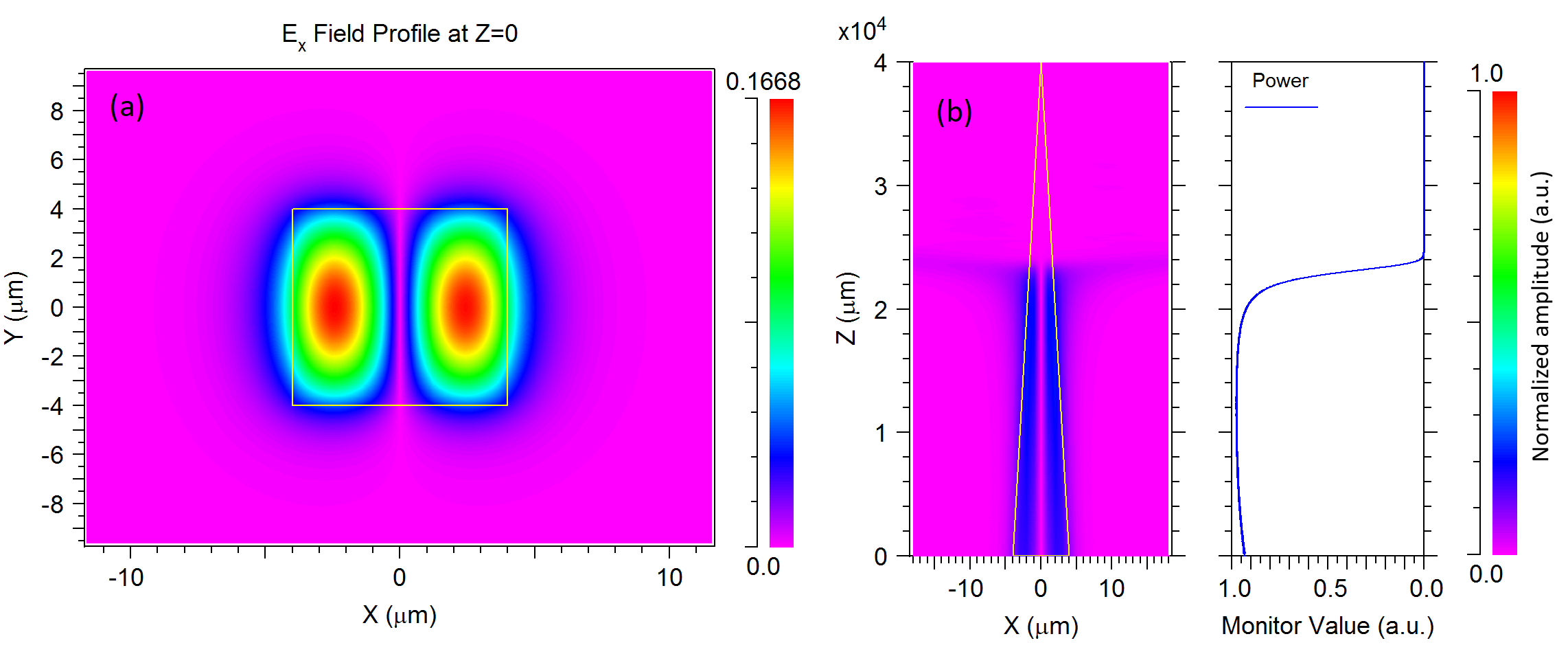}
	\caption{(a) $TEM_{10}$ mode of the symmetric $8\,\upmu \text{m} \times 8\,\upmu$m rectangular channel waveguide core at $1500$ nm. (b) Propagation of the $TEM_{10}$ mode through a $40$ mm long 
square-profile linear taper waveguide. Power monitor indicates loss of confinement.}
	\label{fig:wg_modeprop}
\end{figure}
The low-contrast, square cross-section channel waveguide was modelled as a rectangular core region of width $w$ and height $h=w$ with a refractive index $n_{\text{core}}=1.4738$ embedded in a cladding 
with a with a refractive index $n_{\text{clad}}=1.444$ at $1550$ nm (silica). A mesh resolution of $100$ nm and propagation step size of $2\,\upmu$m was used. The 3D-waveguide was simulated in full-vectorial mode. 
Starting from a multi-mode cross section $w\times h$ of $8\,\upmu\text{m}\times 8\,\upmu$m, the first higher propagating mode was calculated using a BPM mode solver with a two-pass correlation method, 
described in \cite{Feit:80}.
The single mode cut-off core cross-section was determined by simulated adiabatic deconfinement of the $TEM_{01}$ mode, whereby the previously calculated mode was used as the launch field, which 
was propagated through a long, inversely tapered waveguide. The waveguide core parameter $w$ was varied along a linear gradient between $8\,\upmu$m at $z=0\,\upmu$m and $0\, \upmu$m at $z=L$. The 
slow reduction of the waveguide core cross-section along the propagation axis gradually weakens the confinement of the $TEM_{01}$ mode, until the cross-section reaches a critical value at which the waveguide supports 
only the fundamental propagating mode. The waveguide construction used in the simulation is shown in Figure \ref{fig:wg_modeprop}.
\begin{figure}[!ht]
	\centering
	\includegraphics[width=130mm]{ 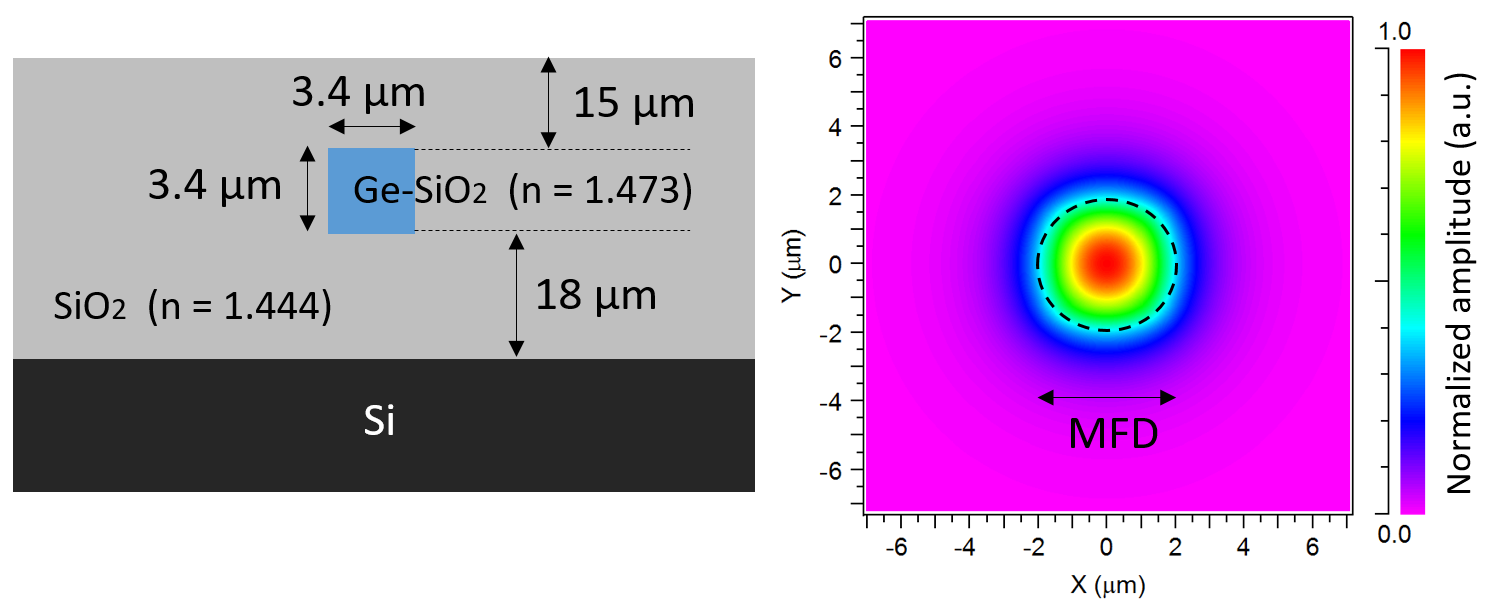}
	\caption{Left: lithographic structure of the waveguide core design (end-on view, not to scale). Right: propagating fundamental waveguide mode of a $3.4\,\upmu\text{m}\times 3.4\,\upmu$m waveguide core at $1500$ nm}
	\label{fig:litho_struct}
\end{figure}
To ensure adiabatic conditions during the simulated mode propagation, a taper length of $40$ mm, correspondingly tapering angle of $10^{-4}$ rad was chosen, satisfying the adiabaticity condition for the local 
angle of the taper wall with the central axis of the waveguide $\theta(z)<\lambda_0/(2w(z) n_{\text{eff}})$ \cite{Burns:77}, which, in our case, suggests a tapering angle $\leq 6\times 10^{-2}$ rad at $z=0$.
The relevant quantity to be determined is the waveguide core dimension $w_0$ at which the $TEM_{10}$ mode of the rectangular waveguide loses confinement and dissipates. Due to the symmetry of the waveguide, the same 
behaviour is expected for the $TEM_{01}$ mode. In this work, the critical waveguide core dimension was determined by observing the propagation of the  mode through the tapered waveguide at $1500$ nm. 
The confinement of the mode field in the waveguide core region during propagation was monitored by integrating the power density in a $8\,\upmu \text{m} \times 8\,\upmu$m rectangle containing the waveguide core at the centre. 
Figure \ref{fig:wg_modeprop} shows the launch mode field used for excitation of the waveguide and the x-z slice of the field propagation through the waveguide.
Loss of mode confinement is indicated by a sudden drop of partial power in the region of measurement around the waveguide core at a critical z-value, defined by the half-maximum location of the monitor curve, beyond 
which point the $TEM_{10}$ mode cannot propagate. In the given case of $\Delta=0.02$ and the aforementioned simulation setup, loss of confinement was observed at $z=2.3\times 10^4\,\upmu$m, or $w=3.4\,\upmu$m. 
Therefore, a waveguide core layer thickness $t=3.4\,\upmu$m and base waveguide width $w=3.4\,\upmu$m were selected for the AWG designs presented in this work. The dissipation of higher modes was confirmed by 
simulations of propagation through a $10\,\text{mm}$ long waveguide. The choice of the largest core cross-section that guarantees single-mode operation at the shortest wavelength results in optimal confinement of the propagating 
mode, which in turn results in less evanescent field interaction between nearest neighbour waveguides, as well as lower bend leakage and thus smaller bend radii, contributing to a higher waveguide integration density.
The method of adiabatic deconfinement proved to be more reliable than correlation-method BPM mode solving. In cases close to the cut-off condition, where multimode propagation was still observed in simulation runs, the 
mode solver failed to find the higher mode, falsely indicating single mode behaviour.The adiabatic inverse tapering method guaranteed the single-mode condition in the simulation and provided a faster way to determine the critical 
core cross-section, requiring only one propagation run instead of multiple runs needed with a mode solving approach.
The schematic cross-section of the waveguide structure and the fundamental propagating waveguide mode are shown in Figure \ref{fig:litho_struct}.

\subsection{Minimal radius of a circular bend}
In commonly used AWG designs, the waveguide array contains curved segments of either fixed or variable curvature. Optimization of the AWG device foot-print affects the radii of curvature in the waveguide array and 
therefore requires prior knowledge of the minimal acceptable waveguide bend radius. In order to study bend leakage in a circular bend waveguide, we have numerically simulated mode propagation through a $10$ mm long circular 
bend using BPM. A series of simulations was performed to study mode propagation through bends of varying radii in a wavelength range between $1500$ nm and $1700$ nm. Mode leakage was measured by monitoring the 
exponential decay of mode power in the waveguide core along the propagation direction. Finally, the bend losses were obtained by linear fitting of the logarithm of the power monitor evolution curves. 
\begin{figure}[!ht]
	\centering
	\includegraphics[width=90mm]{ 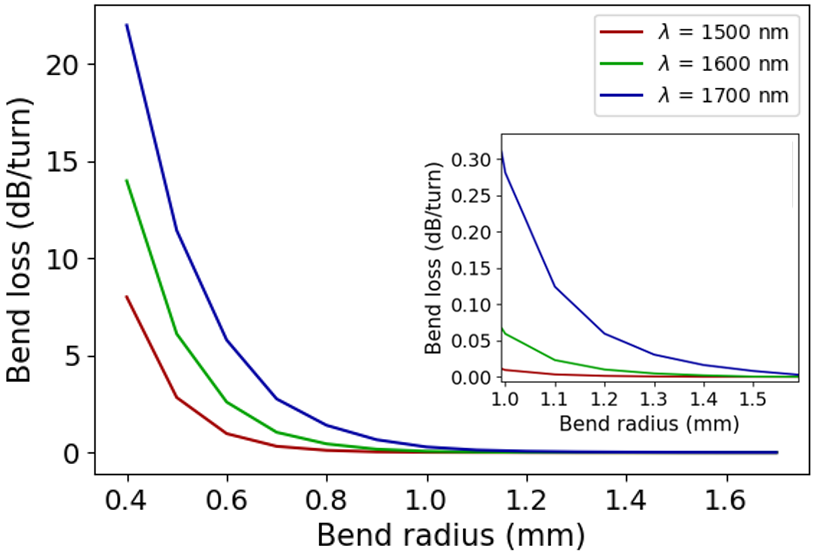}
	\caption{Simulated bend loss for a $3.4\,\upmu \text{m} \times 3.4\, \upmu$m square-profile waveguide core of varying bend radius at three wavelengths in the range $1500$ nm - $1700$ nm.}
	\label{fig:wg_bend}
\end{figure}
Figure \ref{fig:wg_bend} shows the simulated bend losses in a range of bend radii $R_b$ between $0.4$ mm and $1.7$ mm. The results show an increase of bend losses with increasing wavelength for all bend radii in the 
simulated range. Waveguide bends with $R_b<0.7$ mm exhibit prohibitively high leakage of $>1$ dB per $360^{\circ}$ turn  at $1500$ nm and $>3\,\text{dB/turn}$ at $1700$ nm. The losses rapidly decrease with 
increasing bend radius and fall below $10^{-2} \, \text{dB/turn}$ for $R_b>1.4$ mm at $1700$ nm. In this work, a minimum bend radius of $1.5$ mm was chosen in order to reduce bend leakage loss to a 
negligible level between $1.95\times 10^{-6} \, \text{dB/turn}$ at $1500$ nm and $7.98\times 10^{-3} \, \text{dB/turn}$ at $1700$ nm. 

\section{Definition of spectral requirements}\label{sec:spec_req}
The AWG designs constructed in this work are aimed at applications as a core dispersive element in a compact astronomical spectrograph operating in the astronomical near-infrared H-band window 
of atmospheric transparency. The goal of the design is to achieve high throughput ($< 3$ dB insertion loss) as well as a high spectral resolving power. The comparatively small length of the free propagation region on the 
order of 1 cm limits the width of the wavelength-dispersed diffraction image to a few hundreds of $\upmu$m to a few mm. The width of the constructive interference peak on the image plane is fixed by the mode profile of the input 
waveguide and the maximum number of separable spectral lines is severely restricted by the narrow width of the wavelength-dispersed image. Therefore, in order to achieve a high spectral resolution, one has to limit the free spectral 
range of the AWG to a few tens of nm by defining a sufficiently large waveguide array path length increment corresponding to high diffraction orders. In our AWG designs, diffraction orders of up to $m=94$ were used. 
In order to study the feasibility of increasingly higher spectral resolutions, target resolving powers of 15,000, 30,000 and 60,000 (corresponding to spectral resolutions of $0.1$ nm, $0.052$ nm and $0.026$ nm, respectively) 
were defined for low-resolution, mid-resolution and high-resolution designs. Target $FSR$s of $16$ nm, $23$ nm, $32$ nm and $48$ nm were defined to study the performance of waveguide arrays in a range of varying spectral 
bandwidths.

\section{Design of the array-FPR interface}
The starting point of the AWG construction in this work is the design of the array-FPR interface and the fan-out section of the AWG, as it determines the minimum required distance between waveguides at the FPR interface, 
and hence the waveguide width as well as the number of waveguides in the array.

\subsection{Minimum waveguide separation distance}\label{sec:fanout}
\begin{figure}[!ht]
	\centering
	\includegraphics[width=130mm]{ 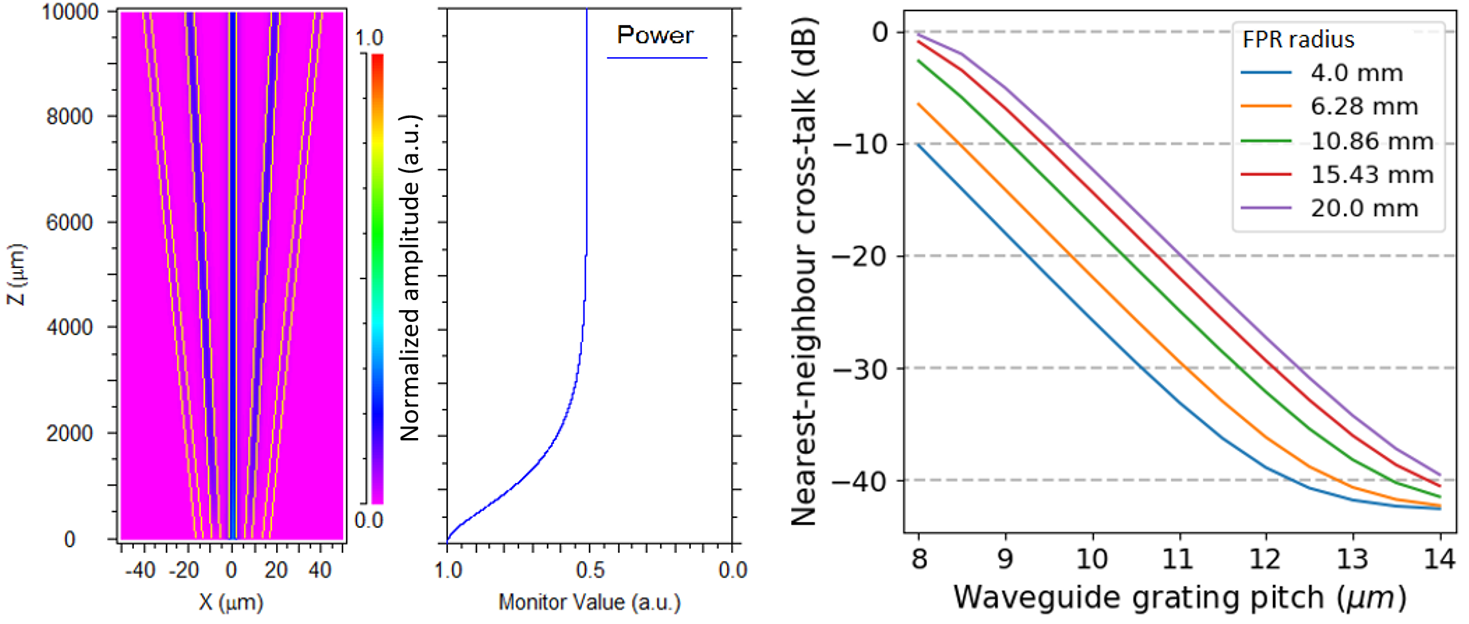}
	\caption{Left: Simplified model of the fan-out region showing propagation of light through a single waveguide at the centre. The power monitor curve shows the transfer of power from the active waveguide to adjacent 
	waveguides. Right: Nearest neighbour cross-talk caused by evanescent field coupling for varying values of the FPR radius, or correspondingly varying focal lengths. Simulated at $\lambda = 1700$ nm.}
	\label{fig:Fanout_crosstalk}
\end{figure}

The integration density of an AWG device partly depends on the maximum amount of acceptable inter-waveguide crosstalk by evanescent field coupling. Depending on the refractive index contrast of the material, a minimum 
distance between waveguides must be observed to prevent degradation of the point spread function due to excessive cross-coupling. The starting point of the AWG design, after definition of the bulk material properties, is the 
determination of the minimum nearest-neighbour waveguide separation at the interface between the waveguide array and FPR slab. In this work, this procedure was performed by numerical simulation of light propagation through 
the fan-out segment of the array-FPR interface using 3D BPM. For this purpose, a simplified model of the fan-out was created in the RSoft CAD utility, shown in Figure \ref{fig:Fanout_crosstalk} (left). Evanescent coupling was 
studied by selectively exciting a single waveguide and observing the transmitted power via a power monitor located inside the waveguide cores. The FPR radius was varied between $4$ mm and $20$ mm and the distance 
between adjacent waveguides at the FPR interface (grating pitch) $d$ was varied between $8\,\upmu$m and $15\,\upmu$m in a series of simulations at $\lambda=1700$ nm, where evanescent field coupling is expected to be 
strongest in the wavelength range of the AWG. The total amount of power transferred to the neighbouring waveguides was defined as $1-P$, where $P$ is the normalized mode power remaining in the initially excited waveguide 
after propagation through the fan-out. 
Figure \ref{fig:Fanout_crosstalk} (right) shows the simulated cross-talk between the waveguides as $10\log(1-P)$ over the grating pitch for a range of grating radii between $R = 4$ mm and $R = 20$ mm. Increasing cross-talk 
can be observed for smaller values of the grating pitch and larger grating radii, which is to be expected since both parameters reduce the fan-out angle of the waveguides and thereby prolongs evanescent interaction, causing 
more power to be transferred between nearest neighbours. The cross-talk in our simulation varies from nearly 0 dB (complete power transfer to nearest neighbours) at $d=8\,\upmu$m and $R = 20$ mm to less than $-40$ dB 
(negligible coupling) at $d=14\,\upmu$m and $R = 4$ mm. In this work, we use $d = 10.57\,\upmu$m as a compromise between waveguide array cross-talk and device foot-print. In our expected range of FPR radii 
used for the AWG designs, the simulation results predict a cross-talk level of $-16$ dB...$-30$ dB in the worst case $\lambda=1700$ nm.  

\subsection{Tapering of the array-FPR interface}\label{sec:WGA-FPR}
\begin{figure}[!ht]
	\centering
	\includegraphics[width=132mm]{ 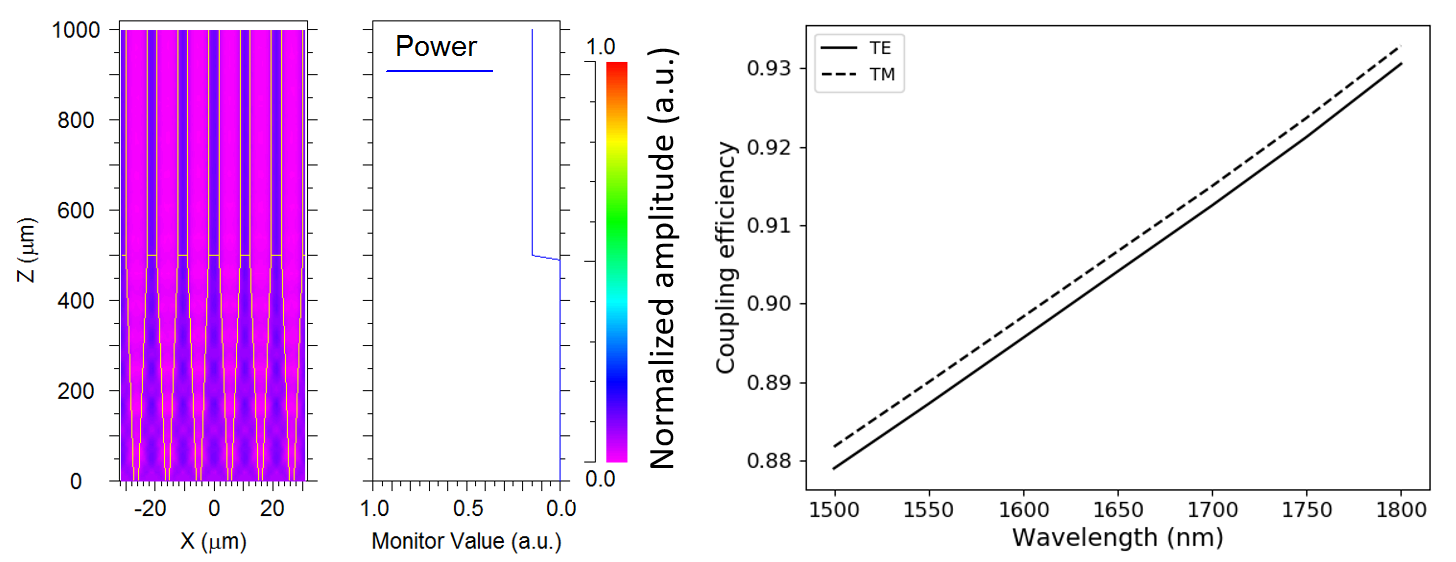}
	\caption{Left: Simulation setup containing five waveguides. Monochromatic continuous-wave field distribution is colour-coded. The pathway monitor shows the normalized power received by the centremost waveguide 
relative to the launch power. The taper segments have a length of $500\,\upmu$m and a base width of $9.07\,\upmu$m. Right: Simulated coupling efficiency of the array-FPR interface for TE and TM-polarized input in the range 
	$1500$ nm - $1800$ nm.}
	\label{fig:WGA-FPR-efficiency-sim}
\end{figure}

Adiabatic mode size conversion plays an important role in integrated photonic circuits. Its primary purpose is the optimization of transmission efficiency between different components of the circuit, such as channel waveguides and 
slab waveguides. Mode size conversion is typically achieved by introducing a longitudinal waveguide core width gradient between an initial core width and the desired final core width. Such an element is called a waveguide taper. 
As an integral part of AWGs, tapers serve to create a smooth transition between the plurality of narrow-core array waveguides and the free propagation region in order to minimize coupling loss by mode matching of the waveguide 
array and the FPR slab waveguide. Since the minimum acceptable waveguide separation distance is significantly larger than the channel waveguide core width, tapering of the array-FPR interface is essential to ensure a high 
transition efficiency. Various types of taper geometries exist, with the most common being the linear horn taper, which is implemented in the AWG designs in this work. A sufficient taper length must be observed to ensure 
quasi-adiabatic mode size conversion, avoiding higher order mode excitation in the multi-mode segment of the taper. 
The light coupling behaviour of the tapered waveguide array was studied numerically using 3D BPM on a simple model of the structure consisting of a few waveguides in an arrangement that resembles the real arrangement of 
waveguides in an AWG device. This approach allows to accurately simulate the isolated coupling efficiency of the array-FPR interface while conserving a substantial amount of computational resources. The BPM simulations were 
performed on a laterally truncated segment of the array-FPR interface, as shown in Figure \ref{fig:WGA-FPR-efficiency-sim}. The array of tapers was excited with a laterally uniform optical field, which was defined as 
$E(x,y)=\psi(y)$, where $\psi(y)$ is the slab waveguide mode function. In order to minimize boundary artifacts, all measurements were taken on the centremost waveguide of the array. The propagation was simulated 
at various wavelengths between $1500$ nm and $1800$ nm. Coupling efficiency was quantified by a mode power monitor situated in the waveguide of interest. The waveguides were arranged equidistantly with a period of 
$10.57\,\upmu$m. A gap of $1.5\,\upmu$m between the edges of the tapers was set in accordance with the safe lower limit of the lithographic process resolution, called minimum feature size, suggested by the integrated 
photonics foundry which implemented the designs into working prototypes. The array of taper waveguides was excited with a uniform slab mode which was normalized to unit power. Since the simulation region contains six 
waveguides, each waveguide in the array received one sixth of the total transmitted power. The efficiency of the interface was calculated as the ratio of the total transmitted power and the input power. The simulation results 
for TE and TM-polarized input are shown in Figure \ref{fig:WGA-FPR-efficiency-sim}. 
The left side of the figure shows the simulation setup after one propagation run. The field amplitude of a monochromatic continuous-wave signal in the simulation region is colour coded in the normalized range 0 - 1. The power 
monitor value is recorded after propagation through the taper. A wavelength-dependent coupling efficiency varying between 0.88 at $1500$ nm and 0.93 at $1800$ nm is observed for TE-polarized input. The coupling efficiency 
is marginally higher for the TM polarization. A study of taper efficiency in an earlier AWG design on a $\Delta=0.015$ silica-on-silicon platform was presented in \cite{Stoll:17}. In this study, a $1$ mm long linear taper of 
$14\,\upmu$m base width was found to exhibit a simulated coupling loss of $0.31$ dB at $1500$ nm. The coupling loss was reduced to 0.088 dB by introducing a novel discontinuous taper structure which takes advantage of 
multi-mode interference. The MMI taper is one of the possible loss reduction mechanisms under consideration for future iterations of the AWG designs.

\subsection{Aperture of the waveguide grating}\label{sec:grating_aperture}
In order to achieve a high transmission efficiency as well as minimize truncation sidelobes of the diffraction image, it is important to design a waveguide grating with a sufficiently large aperture, i.e. angular width of 
the array-FPR interface, which covers most of the input beam cross-section. 
\begin{figure}[!ht]
	\centering
	\includegraphics[width=130mm]{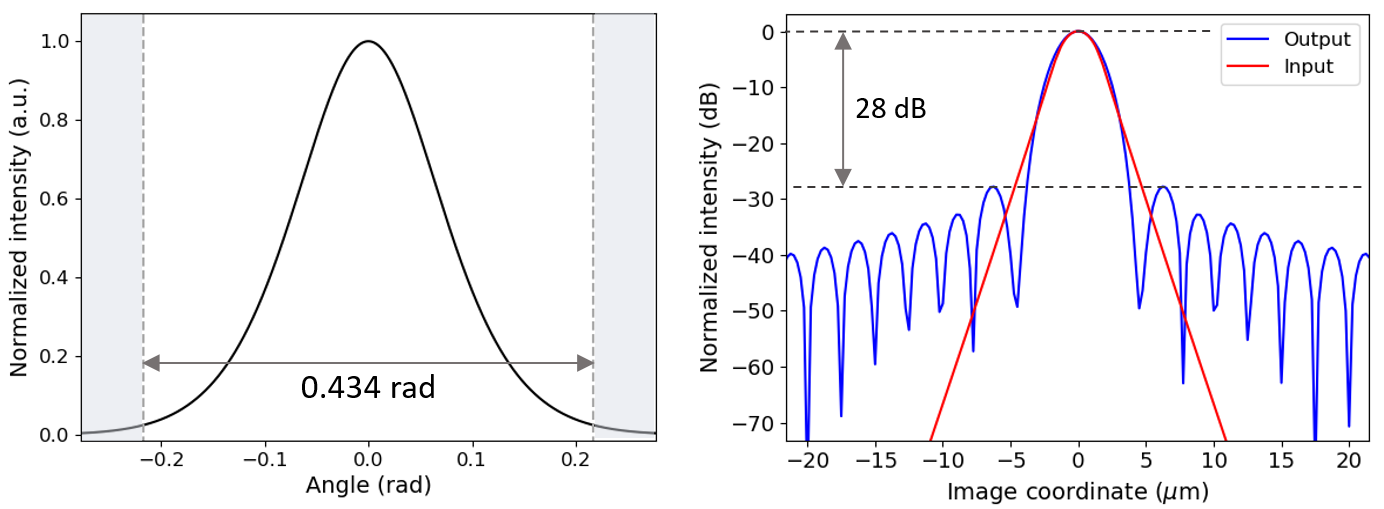}
	\caption{Left: Normalized intensity distribution of the input waveguide far field with truncation boundaries indicated by dashed lines. Right: Reconstructed near field intensity profile of the truncated far field 
distribution (blue), input field intensity profile (red).}
	\label{fig:input_farfield}
\end{figure}
On the other hand, the aperture size should be as small as possible in order to minimize the foot-print and number of waveguides of the AWG structure, reducing the potential for phase errors. The calculation of the 
waveguide array aperture size is based on the known near-field distribution of the input waveguide mode, from which an angular far-field distribution at the FPR-array junction is derived. In this work, we have designed 
the width of the waveguide grating to cover $99\%$ of the incoming beam power, corresponding to a negligible truncation loss $0.044$ dB. The associated aperture angle covering $99\%$ of the beam power is 0.434 rad.
The left side of Figure \ref{fig:input_farfield} shows the angular intensity profile of the input beam far field. The dashed lines mark the boundaries of the region containing $99\%$ of the input beam power. The right side 
of the figure shows the normalized input intensity distribution (red curve) and the normalized output intensity distribution (blue curve). The truncation of the beam to 0.434 rad induces sidelobes of -28 dB peak intensity in the output 
diffraction image.

The calculation of the minimum required focal length from the minimum waveguide separation distance, angular grating aperture and resolution requirement is described in the next section.

\section{Geometrical construction of the AWG layout}
The structure of the AWG can be arranged in various ways. The choice of an appropriate waveguide layout mainly depends on the grating order and thus the geometrical path length increment of the waveguides. AWG geometries 
can be roughly distinguished into three types: low-order, medium-order and high-order AWG devices, each requiring a specific arrangement of waveguides. Low-order AWGs with small path length increments are suitable for wide-band, 
low-resolution wavelength demultiplexing. The waveguide array of this AWG type is arranged in an S-geometry, which allows for very small path length increments while maintaining the required minimum distance between waveguides. 
The S-type is suitable for large free spectral ranges above $50$ nm. Medium-order AWGs are typically realized in the conventional horse-shoe geometry. This type of AWGs has been widely used for DWDM applications in optical 
telecommunications, as they offer a good balance between spectral resolution and spectral bandwidth with free spectral ranges between $30$ nm and $50$ nm in this particular application case. High-order AWGs with a large path 
length increment can achieve high spectral resolutions at the cost of reduced free spectral range. This type of waveguide array requires sufficient space for path length accumulation. Therefore, a folded design is chosen which allows 
the waveguides to "balloon" outwards. The free spectral range of this AWG type is typically $<20$ nm. In this work, horse-shoe-type and folded-type AWGs are selected and implemented for a wide FSR at low-medium spectral 
resolutions and a narrow FSR at low-high spectral resolutions, respectively.

This work presents five different AWG devices designed to operate in spectral orders between $m = 31$ and $m = 94$. Two AWG designs were implemented as folded-type AWGs, from here on denoted as type A and type B. 
Three designs were implemented as horse-shoe type AWGs, referred to as type C, D and E. Designs of type A and B operate in the same spectral order $m = 94$, which corresponds to a $FSR$ of $16$ nm. Design A is the largest 
AWG in the group, having a focal length of $18$ mm and a waveguide array size of 722. Design B is directly derived from design A through down-scaling by a factor of 0.5. Since the input waveguide properties are identical in all six 
AWG variants, they all share the same angular grating aperture of 24.4 degrees. The AWG layouts were implemented with careful observation of geometrical consistence, avoiding any discontinuities of optical paths, maintaining the 
minimum waveguide separation distance, minimum lithographic feature size, as well as a minimum waveguide bend radius to minimize bend leakage. All designs were equipped with 30 output waveguides evenly distributed over the 
output facet of the slab waveguide for characterization purposes.

\subsection{Calculation of FPR length based on resolution criterion}
Given a fixed size of the input waveguide core, the spectral resolution of the AWG is determined by the waveguide array path length increment, the number of waveguides, the grating pitch and the focal length, which is identical to 
the length of the FPR slab waveguide. Requiring a certain spectral resolution $\Delta \lambda$, one must ensure that overlapping peaks corresponding to $\lambda$ and $\lambda+\Delta \lambda$ can be identified as separate features 
in the spectral image. In this work, we obtain lower estimates for the required focal length using the full-width-half-maximum (FWHM) resolution criterion, assuming two spectral lines separated by approximately one FWHM 
are just barely resolvable as separate intensity maxima. The necessary condition for the focal length is derived from the AWG grating equation \cite{Takahashi:95}, yielding a change of the diffraction angle $\theta$ at $\lambda_0$ as
\begin{equation}
	\Delta \theta \approx \left.\frac{d\theta}{d\lambda}\right|_{\lambda_0}\Delta\lambda = \frac{\lambda_0^2}{R\cdot FSR \cdot d \cdot n_s}
\end{equation} 
where $n_s$ is the slab waveguide effective index, $d$ is the grating pitch and $\Delta \lambda \ll FSR$. Requiring $L_f \Delta\theta \geq w_{1/2}$, we obtain the condition for the focal length $L_f$ for any given combination of 
spectral resolving power $R$ and $FSR$ as
\begin{equation}
	L_f \geq \frac{R\cdot FSR \cdot n_s \cdot d}{\lambda_0^2}w_{1/2},
\end{equation} 
where $w_{1/2}$ is the $FWHM$ of the squared input waveguide mode field.
\begin{table}[!ht]
	\caption{\label{tab:min_Lf} Minimum required focal length in mm as a function of free spectral range and resolving power.}
	\centering
	\begin{tabular}{c| c c c c c c c} 
	 	\hline
		 \backslashbox{R}{FSR (nm)}& 16 & 23 & 32 & 48\\ [0.5ex] 
		 \hline
		 15,000 & 4.055 & 5.83 & 8.11 & 12.165  \\ 
		 30,000 & 8.11 & 11.66 & 16.22 & 24.33  \\
		 60,000 & 16.22 & 23.32 & 32.44 & 48.66 \\
		 \hline
	\end{tabular}
\end{table}
The input waveguide with a $2\%$ refractive index contrast core measuring $3.4\,\upmu \text{m} \times 3.4\,\upmu$m produces a mode field with an intensity $FWHM$ of $2.62\,\upmu$m and is used in all AWG designs presented 
in this work. The slab waveguide effective index at $1550$ nm is calculated as $n_s = 1.4661$. Given $d=10.57\,\upmu$m and $\lambda_0=1550$ nm, we obtain the minimum required focal length as a function of $FSR$ and 
$R$ only, presented in Table \ref{tab:min_Lf}.

\subsection{Construction of the waveguide array}
As the central element of the AWG device, the waveguide array serves as a multipath optical delay line consisting of several hundreds of single-mode waveguides of length $L_j = L_{j-1}+\Delta L$, where $j$ is the waveguide number 
and $\Delta L=m\lambda_0/n_{\text{eff}}$ is a fixed path length difference between nearest neighbour paths, defined by the integer grating order $m$, the central wavelength $\lambda_0$ and the waveguide effective index 
$n_{\text{eff}}$. Since the spectral requirements imposed on the designs in section \ref{sec:spec_req} define $\lambda_0$ and the free spectral range as the design parameters, the appropriate grating order $m$ must be determined 
for each design. The free spectral range of the AWG is given by $FSR=n_{\text{eff}}n_g\lambda_0/m$, where $n_g$ is the waveguide group index. Hence, the grating order is calculated as
\begin{equation}
	m = \left[\frac{n_{\text{eff}}\lambda_0}{n_g FSR}\right],
\end{equation}
where the square brackets on the right hand side indicate rounding to the nearest integer. The values $n_{\text{eff}}$ and $n_g$ were obtained from numerical simulations of the array waveguide core as $n_{\text{eff}}=1.4586$ 
and $n_g=1.4959$ at $\lambda_0=1550$ nm.
\begin{figure}[!ht]
	\centering
	\includegraphics[width=70mm]{ 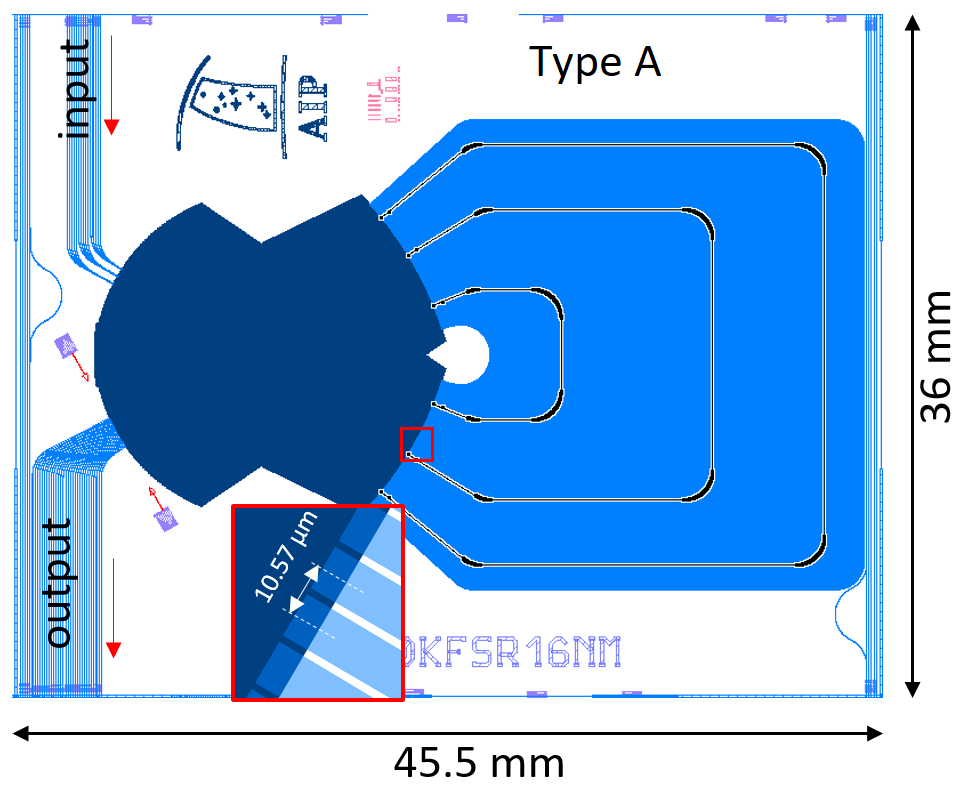}
	\caption{Lithographic mask layout of the largest AWG design (Type A). Theoretical $R=60,000$. Three selected array waveguides highlighted showing straight and curved waveguide segments.}
	\label{fig:AWG_Type_A}
\end{figure}
\begin{figure}[!ht]
	\centering
	\includegraphics[width=125mm]{ 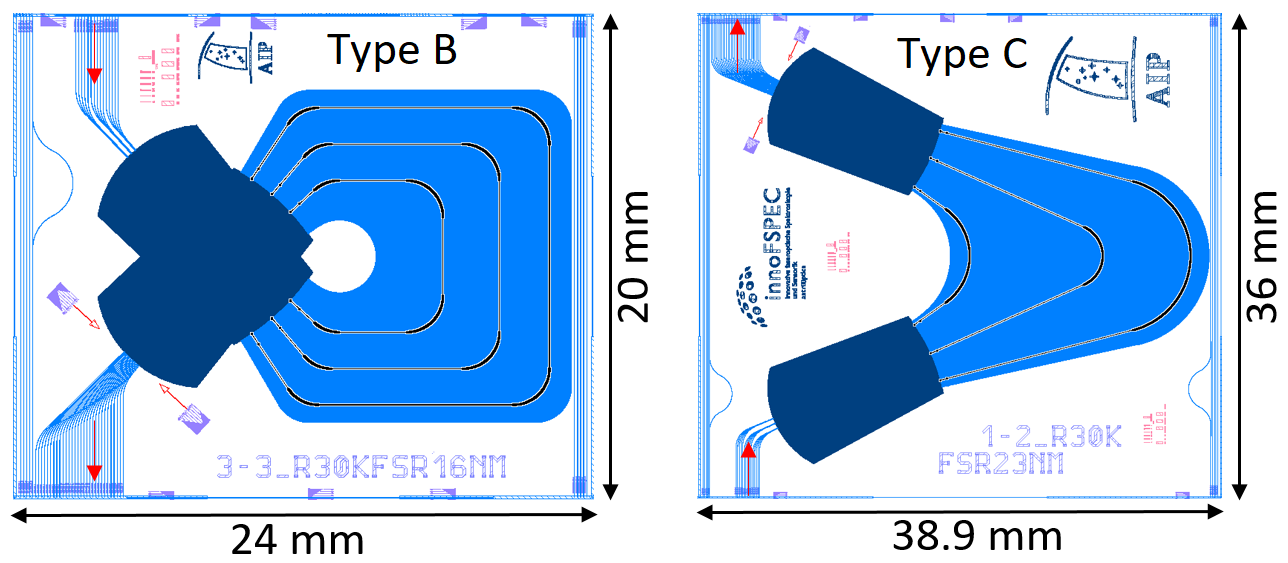}
	\caption{Lithographic mask layouts of AWGs B and C. Theoretical $R=30,000$. Type B is conceptually similar to type A, containing five straight segments and four curved segments in the waveguide array. Type C is a traditional design 
	consisting of two straight segments and one curved segment.}
	\label{fig:AWG_Type_B_C}
\end{figure}
\begin{figure}[!ht]
	\centering
	\includegraphics[width=125mm]{ 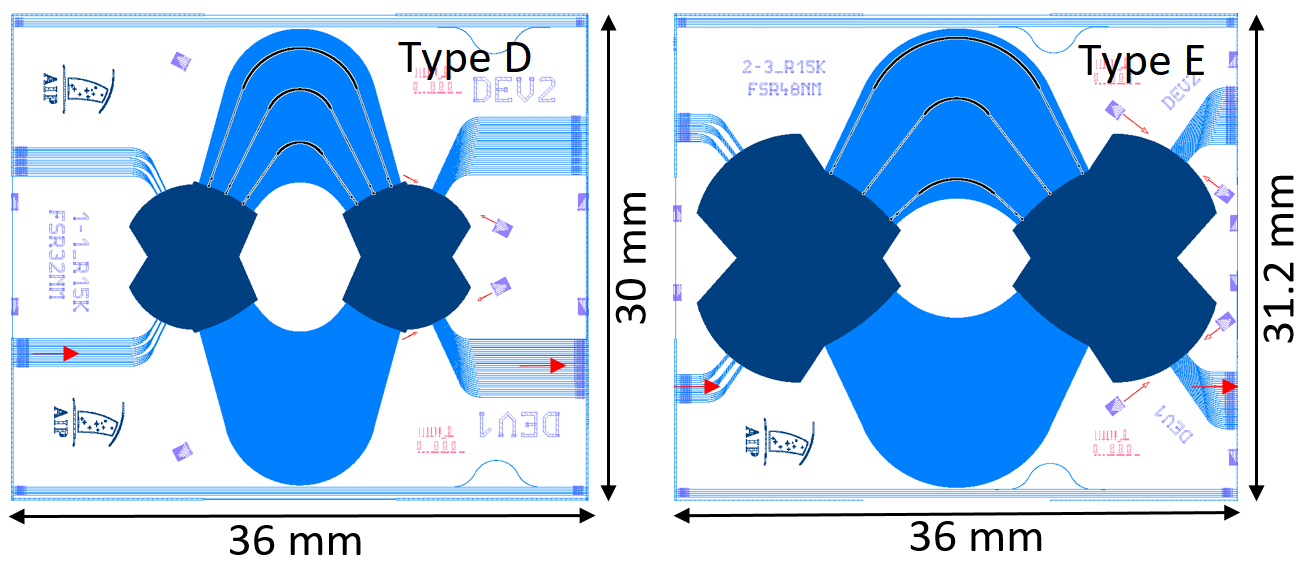}
	\caption{Lithographic mask layouts of AWGs D and E. Theoretical R=15,000. Both AWGs are conceptually similar to type C.}
	\label{fig:AWG_Type_D_E}
\end{figure}
\begin{table}
	\caption{\label{tab:AWG_specs} Specifications of AWG designs A-E.}
	\centering
	\begin{tabular}{c c c c c c c c} 
	 	\hline
		 Type & FSR (nm) & $\Delta \lambda$ (nm) & R & $L_f$ (mm) & $N_{wg}$ & $\Delta L$ ($\upmu$m) & Footprint ($\text{cm}^2$)\\ [0.5ex] 
		 \hline
		 A & 16.08 & 0.026 & 60,000 & 18 & 722 & 99.89 & 16.38 \\ 
		 B & 16.08 & 0.052 & 30,000 & 9 & 360 & 99.89 & 4.8 \\
		 C & 22.9 & 0.052 & 30,000 & 12.728 & 512 & 70.1 & 14.0\\
		 D & 32.2 & 0.1 & 15,000 & 9 & 362 &  49.9 & 10.8 \\
		 E & 48.8 & 0.1 & 15,000 & 13.5 & 543 & 32.9 & 11.23 \\ [1ex] 
		 \hline
	\end{tabular}
\end{table}

The required grating order values were calculated as $m=94$, $m=66$, $m=47$ and $m=31$, corresponding to free spectral ranges of $16.08$ nm, $22.9$ nm, $32.2$ nm and $48.8$ nm, respectively. Note that, for any given central 
wavelength, the FSR is restricted to a discrete set of values which corresponds to the integer values of $m$. Therefore, the actual FSR deviates slightly from the design values used to calculate $m$. With the known grating order values, 
the geometrical path length increments $\Delta L$ of the AWG designs was calculated accordingly as $99.89\,\upmu$m, $70.1\,\upmu$m, $49.9\,\upmu$m and $32.9\,\upmu$m. Appropriate waveguide array geometries were selected for 
each case in order to minimize the foot-print of the designs as well as the length of the optical paths. The waveguide arrays with the largest path length increment of $99.89\,\upmu$m were implemented in a folded arrangement consisting 
of nine segments of alternating straight and curved waveguide sections with a constant curvature radius of $1.5$ mm, as shown in Figure \ref{fig:AWG_Type_A}. The distance between nearest-neighbour waveguides in the central 
bulk of the array was required to be $\geq 15\,\upmu$m in order to prevent evanescent field coupling. The folded geometry was used for AWG designs of type A and type B, while the traditional horseshoe geometry was used for 
types C, D and E. The largest design of type A contains 722 waveguides with lengths between $7$ mm and $79$ mm. The AWG design of type B, shown on the left of Figure \ref{fig:AWG_Type_B_C},  is a down-scaled version of 
type A, containing 360 waveguides. AWG designs with of the horseshoe type are shown in Figure \ref{fig:AWG_Type_B_C} (right) and Figure \ref{fig:AWG_Type_D_E}. The array waveguides of these designs consist of two straight sections 
and a central curved section of variable radius $\geq 1.5$ mm. In design C, the bend radius varies from $1.654\,\text{mm}$ to $6.635\,\text{mm}$. The variation in design D is between $1.586\,\text{mm}$ and $4.803\,\text{mm}$ and 
design E uses bends with radii between $2.025\,\text{mm}$ and $6.570\,\text{mm}$. The smallest radii are found roughly in the first third of the waveguide array, counting from the shortest waveguide. Despite an equal resolving power, 
the required number of waveguides is larger in type C as compared to type B due to the larger $FSR$ of $23$ nm. The nearest-neighbour path length increment is $70.1\,\upmu$m in the type C design.
The foot-prints of the final designs range from $4.8\,\text{cm}^2$ (type B) to $16.38\,\text{cm}^2$ (type A). Each design contains arrays of input and output waveguides, which do not use tapering in order to minimize the spot size of 
the input near field and thus maximize the achievable resolution. An array of 31 output waveguides distributed equidistantly across the spectral image region of the output FPR facet was integrated in each design for testing purposes.
The design specifications of all five AWG types are summarized in Table \ref{tab:AWG_specs}.

\section{Numerical simulation of the AWG designs}
We have numerically studied the spectral properties of the AWG designs using a segmented simulation approach which employs the beam propagation method in combination with a scalar method based on Fraunhofer diffraction. 
In the first stage, the optical properties of fundamental AWG building blocks were simulated using 3D-BPM. Specifically, mode profiles of the input waveguides, array waveguides and slab waveguides, as well as their wavelength-dependent 
effective indices were obtained for the wavelength range of operation. The numerical model included the wavelength-dependent refractive index of the $SiO_2$ cladding from the dielectric material library of the RSoft suite, while 
the refractive index of the core was calculated from the refractive index of the cladding assuming a fixed contrast of 0.02. The mode profiles of the input waveguides and array waveguides were further processed to obtain their respective 
far field functions. The pre-calculated far field and effective index data was then used to set up the scalar diffraction model. In the second stage, the wavelength-dependent transmission from input waveguides $j$ to output waveguide $k$ 
was calculated according to \cite{Okamoto:2006, Stoll:20}
\begin{subequations}
	\begin{equation}\label{eq:AWG_analytical}
		T_{jk}(\lambda)=\left|\sum_{l=1}^{N}c_{jk}(l)\exp\left(-i\Omega_{jk}(l)\right)\right|^2
	\end{equation}
	\begin{equation}
		c_{jk}(l) = f(\sigma_{jl})g(\rho_{jl})g(\rho_{kl})h(\sigma_{kl})
	\end{equation}
	\begin{equation}
		\Omega_{jk}(l) = \frac{2\pi}{\lambda}\left( n_s(r_{jl}+r_{kl})+ n_a((l-1)\Delta L+L_1) \right),
	\end{equation}
\end{subequations}
where $\Omega_{jk}$ is the accumulated phase after propagation between input $j$ and output $k$, $f$, $g$ and $h$ are angular far field functions of the input wave\-guides, array wave\-guides and output wave\-guides, respectively, 
$n_a$ and $n_s$ are effective indices of the waveguide array and the FPR slab, respectively, $r_{jl}$ and $r_{kl}$ are the free propagation distances between the terminus of array waveguide $l$ and input waveguide $j$ / output waveguide 
$k$, respectively, $\sigma_{jl}$ and $\rho_{jl}$ are propagation angles relative to the axis of the input waveguide $j$ and array waveguide $l$, respectively, and $\Delta L$ is the waveguide array path length increment.
\begin{figure}[!ht]
	\centering
	\includegraphics[width=80mm]{ 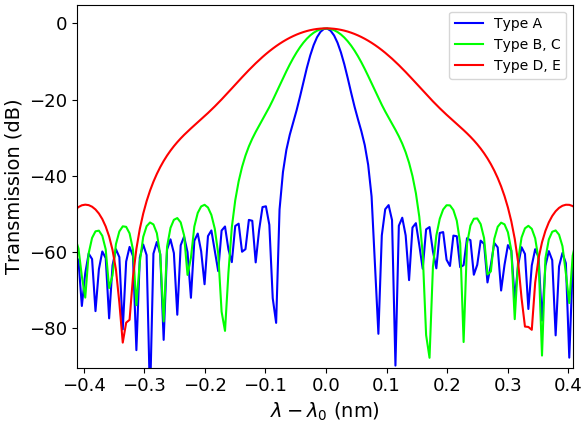}
	\caption{Power transmission curves of the centremost output channel of AWG types A-E.}
	\label{fig:AWG_Type_ABCDE}
\end{figure}
The wavelength transmission spectrum of any given output wave\-guide is calculated by summation over all contributions from N array waveguides. Implementation of different AWG designs in this model was accomplished by variation of the 
path length increment $\Delta L$ and the N geometry parameters $r_{kl}$. 

We have obtained theoretical wavelength channel transmission curves for all five AWG types, calculation results being shown in Figure \ref{fig:AWG_Type_ABCDE}. The transmission includes an insertion loss of $1$ dB, obtained from simulations 
of the array-FPR taper loss and truncation loss in both star couplers, neglecting fibre-chip interfacing loss. Waveguide coupling by evanescent field interaction was not included in this simulation due to the limitations of the AWG model. 
A full simulation using BPM was not feasible due to the size of the AWG structures. Based on the results in section \ref{sec:fanout}, we assumed the coupling and its effects to be negligibly small. The impact of evanescent interaction of array 
waveguides on the performance of high-resolution AWGs is a subject of future investigation. The three distinct transmission curves in Figure \ref{fig:AWG_Type_ABCDE} correspond to the different levels of spectral resolving power. 
The transmission curves are identical for AWGs of equal resolving power, since identical input/output waveguides are used in all designs. The $3$ dB-bandwidths of the transmission curves were found to be $0.034$ nm for the AWG of type A, 
$0.067$ nm for AWGs of types B and C and $0.13$ nm for AWGs of types D and E. These values shall be used as reference in the experimental evaluation of the AWG designs.

\section{Fabrication and characterization}
\begin{figure}[!ht]
	\centering
	\includegraphics[width=100mm]{ 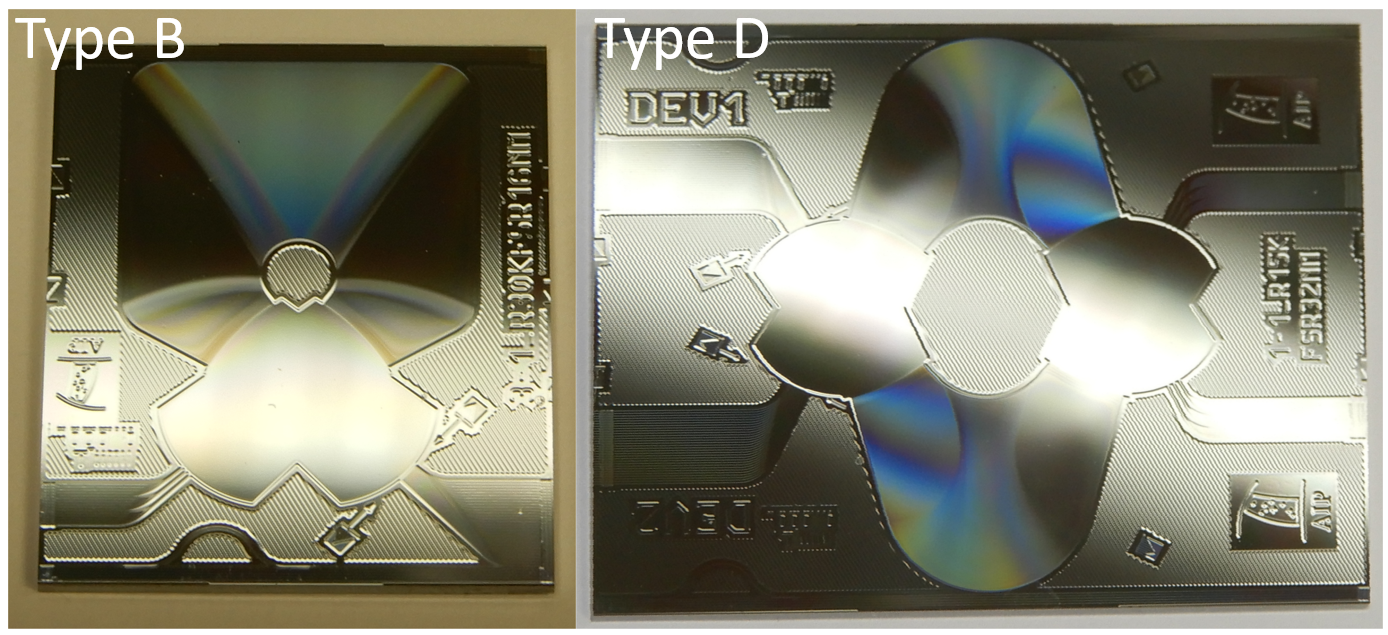}
	\caption{Fabricated AWG devices of folded-type (left) and horse-shoe-type (right).}
	\label{fig:AWG_fab}
\end{figure}

The AWG devices A-E were fabricated by an external foundry (Enablence USA Components, Inc) on a 6-inch Si wafer using UV-photolithography and APCVD technology. A $3.4\, \upmu$m thick Ge-doped silica core with $n_c = 1.473$ 
was deposited on $20\,\upmu$m of thermal oxide, followed by reactive ion etching (RIE) of the waveguide structure and deposition of a $15\,\upmu$m thick upper cladding with $n_{\text{cl}} = 1.444$. Figure \ref{fig:AWG_fab} shows two fabricated 
AWG chips of type B (left) and type D (right), implemented in a folded geometry and horse-shoe geometry, respectively.

A preliminary characterization of the fabricated AWGs was conducted using a T100S-HP CL band tunable laser source (TLS) coupled to a Yokogawa AQ6151 optical wavelength meter (OWM) via a PC running a LabView controller module. Coupling between the TLS and the 
AWG chips was established via SM1500 fibres with a mode field diameter of $4.2\,\upmu$m, matching the on-chip waveguide mode $1/e$ field diameter. The power transmission spectrum was measured at selected output channels to evaluate 
the channel FWHM and device insertion loss. Measurements were taken relative to a straight reference waveguide located on the same chip. 

\subsection{Wavelength channel transmission}\label{sec:trans_measurement}
\begin{figure}[!ht]
	\centering
	\includegraphics[width=80mm]{ 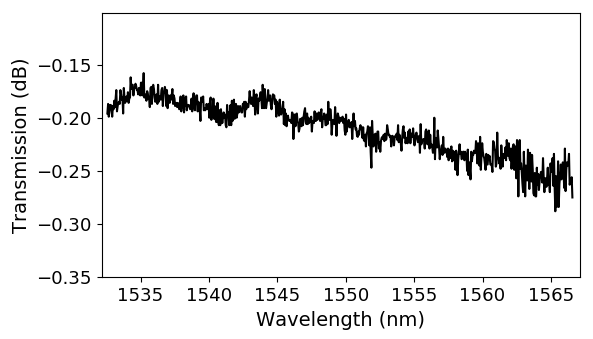}
	\caption{Transmission spectrum of a $36$ mm long reference waveguide relative to TLS background (TE polarization).}
	\label{fig:Meas_REF}
\end{figure}

\begin{figure}[!ht]
	\centering
	\includegraphics[width=130mm]{ 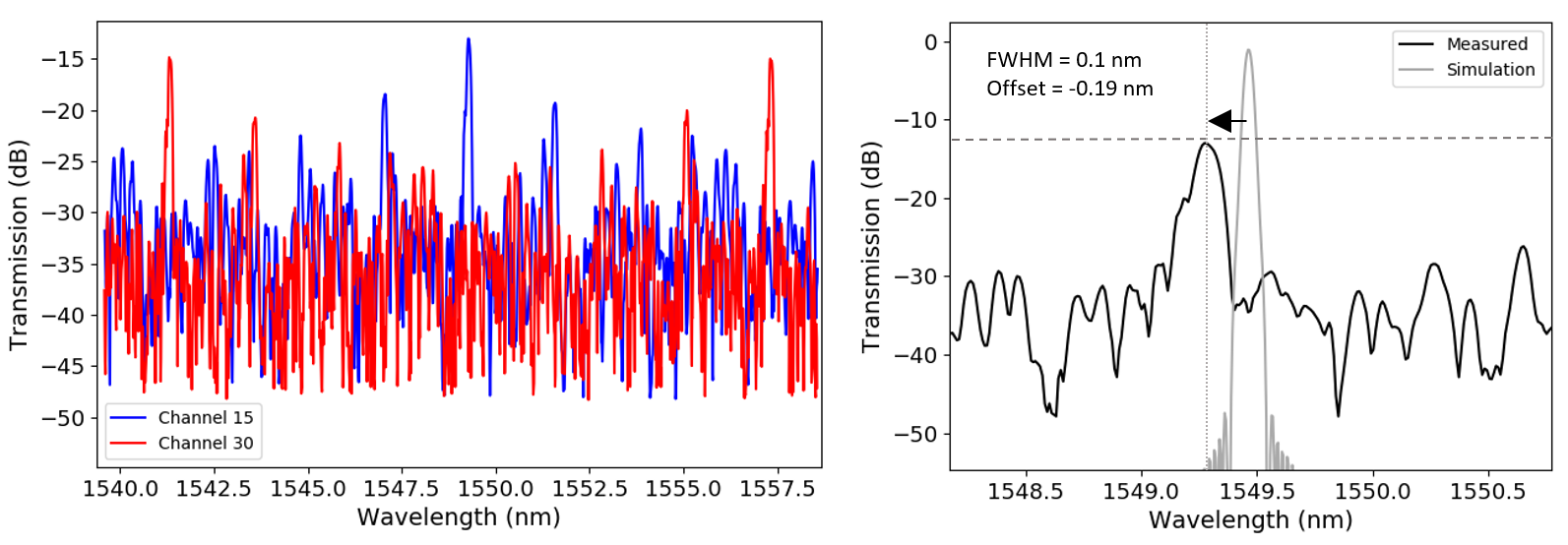}
	\caption{(a) Type A transmission spectrum of output channels 15 and 30. (b) Main transmission peak in comparison with simulated transmission curve. The measured peak is offset from its theoretical wavelength by $-0.17$ nm and its 
3-dB-bandwidth is $0.1$ nm.}
	\label{fig:Meas_TypeA}
\end{figure}

\begin{figure}[!ht]
	\centering
	\includegraphics[width=130mm]{ 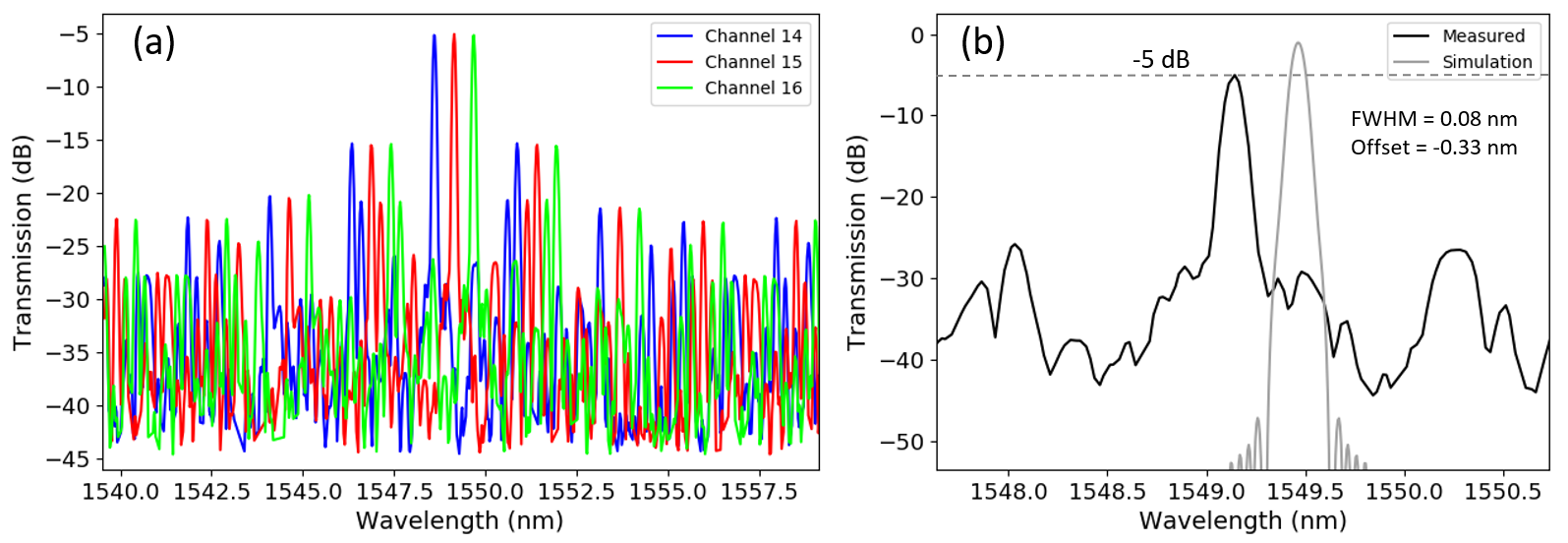}
	\caption{(a) Type B transmission spectrum of output channels 14-16. (b) Main transmission peak in comparison with simulated transmission curve. The measured peak is offset from its theoretical wavelength by $-0.33$ nm and its 
3-dB-bandwidth is $0.082$ nm.}
	\label{fig:Meas_TypeB}
\end{figure}

\begin{figure}[!ht]
	\centering
	\includegraphics[width=130mm]{ 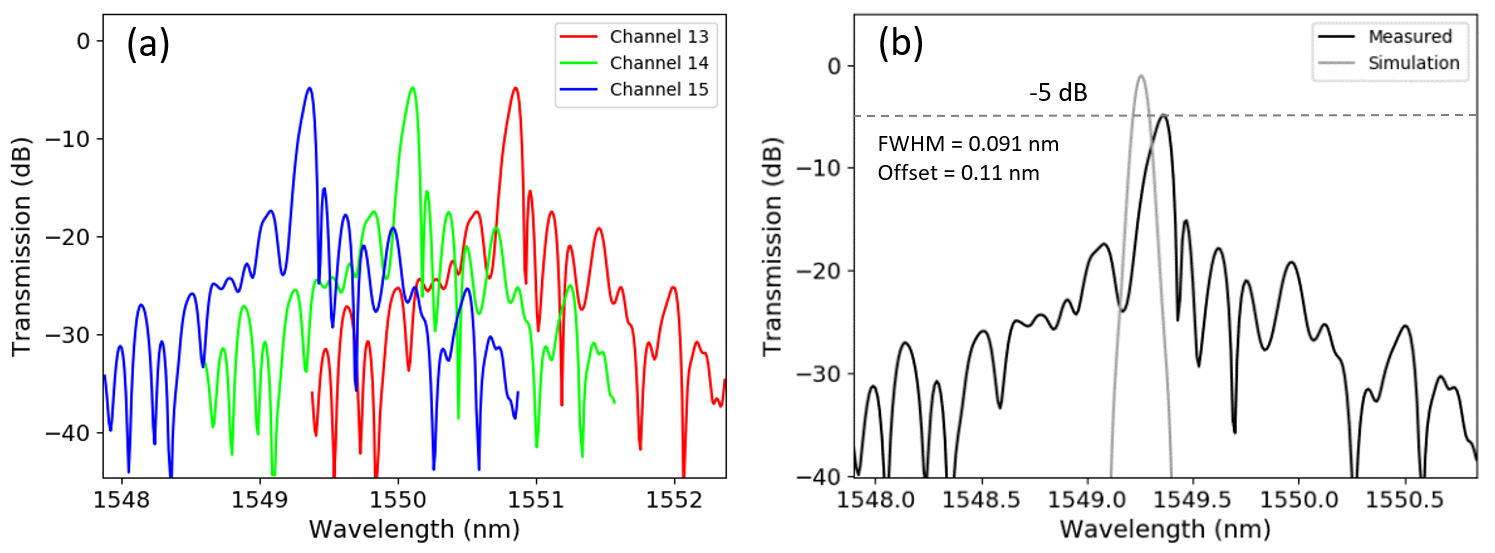}
	\caption{(a) Type C transmission spectrum of output channels 13-15. (b) Main transmission peak in comparison with simulated transmission curve. The measured peak is offset from its theoretical wavelength by $0.09$ nm and its 
3-dB-bandwidth is $0.09$ nm.}
	\label{fig:Meas_TypeC}
\end{figure}

The wavelength transmission spectrum of the fabricated AWGs was measured for selected output waveguides by injection of monochromatic light into the central input waveguide and measurement of the transmitted power on each output 
waveguide. The wavelength of the TLS was tuned in discrete steps of $0.05$ nm. Prior to each set of measurements, reference spectra were recorded for direct fibre-fibre coupling and fibre-waveguide-fibre coupling. Figure \ref{fig:Meas_REF} 
shows the transmission spectrum of a $36$ mm long straight reference waveguide relative to the TE-polarized TLS background. The measured insertion loss of the straight reference waveguide ranges between $0.16$ dB at $1530$ nm  and 
$0.28$ dB at $1570$ nm.
The first AWG under test was the largest design of type A. This design is very susceptible to aberrations of the waveguide effective index due to refractive index non-uniformities and film thickness, as well as waveguide width variations. 
The transmission spectra of the centremost output channels are shown in Figure \ref{fig:Meas_TypeA}. The measured transmission curve is significantly broadened in comparison with the simulated transmission curve, as shown in Figure 
\ref{fig:Meas_TypeA}(b). The strong degradation of spectral properties can be attributed to the large average length of the waveguide array of $40$ mm, leading to unacceptably large phase errors (see \cite{Stoll:17}). In a previous work, 
we have obtained an upper estimate of the effective index variation of $\delta n_{\text{eff}}=7.2\times10^{-6}$ by measuring the variations of optical lengths in a silica AWG from the same batch as the devices in this work \cite{Stoll:20}. 
The estimated spectral resolving power of the fabricated type A device is $R>15,000$ at $1550$ nm. 
The measurement results of the type B AWG are shown in Figure \ref{fig:Meas_TypeB}. Being a downscaled type A design, the AWG shows similar results, albeit with noticeably lower crosstalk level, as should be expected with a reduction of 
the average propagation length. Figure \ref{fig:Meas_TypeB}(a) shows the transmission spectra of three output channels in the central region of the output waveguide interface. The main transmission peak is shown in Figure \ref{fig:Meas_TypeB}(b).
We have estimated a spectral resolving power $R>18,900$ at $1550$ nm for device B. The overall comparison with type A results shows a significant improvement of transmission loss and narrowing of the transmission peak bandwidth. 
Similar measurements were performed on AWGs of types C-E. The measurement results for type C are shown in Figure \ref{fig:Meas_TypeC}. These results are comparable with type B measurements, with the exception of the larger FSR of type C.

\begin{figure}[!ht]
	\centering
	\includegraphics[width=130mm]{ 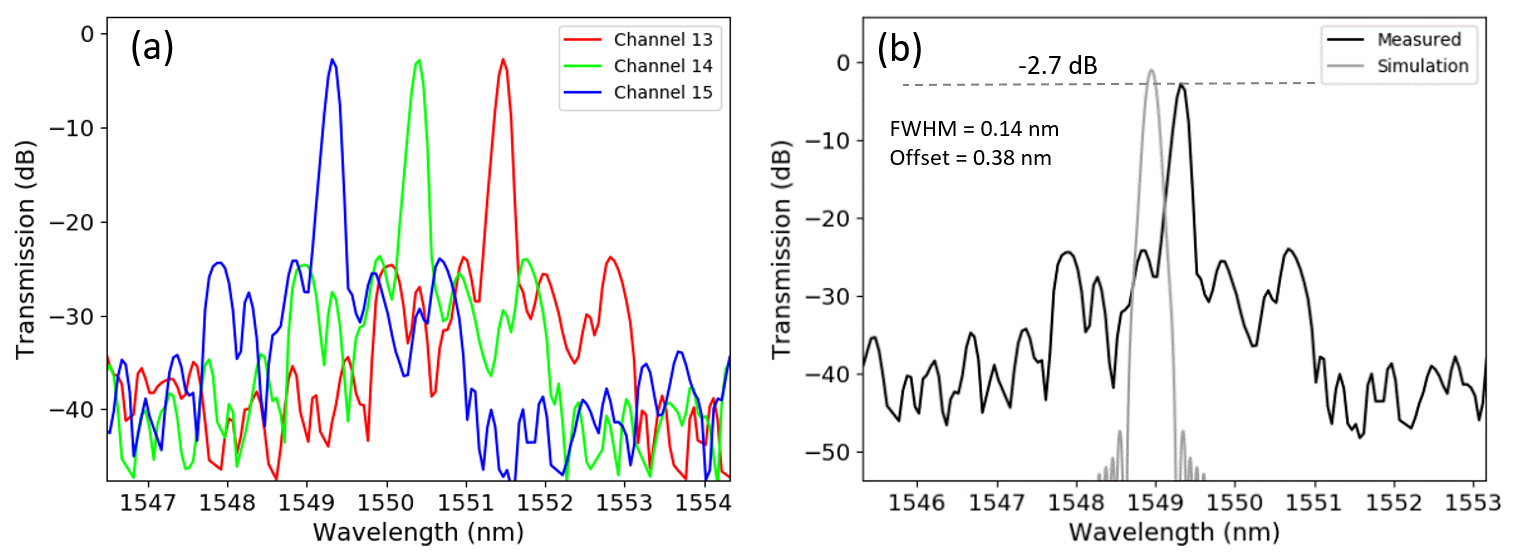}
	\caption{(a) Type D transmission spectrum of output channels 13-15. (b) Main transmission peak in comparison with simulated transmission curve.}
	\label{fig:Meas_TypeD}
\end{figure}

\begin{figure}[!ht]
	\centering
	\includegraphics[width=130mm]{ 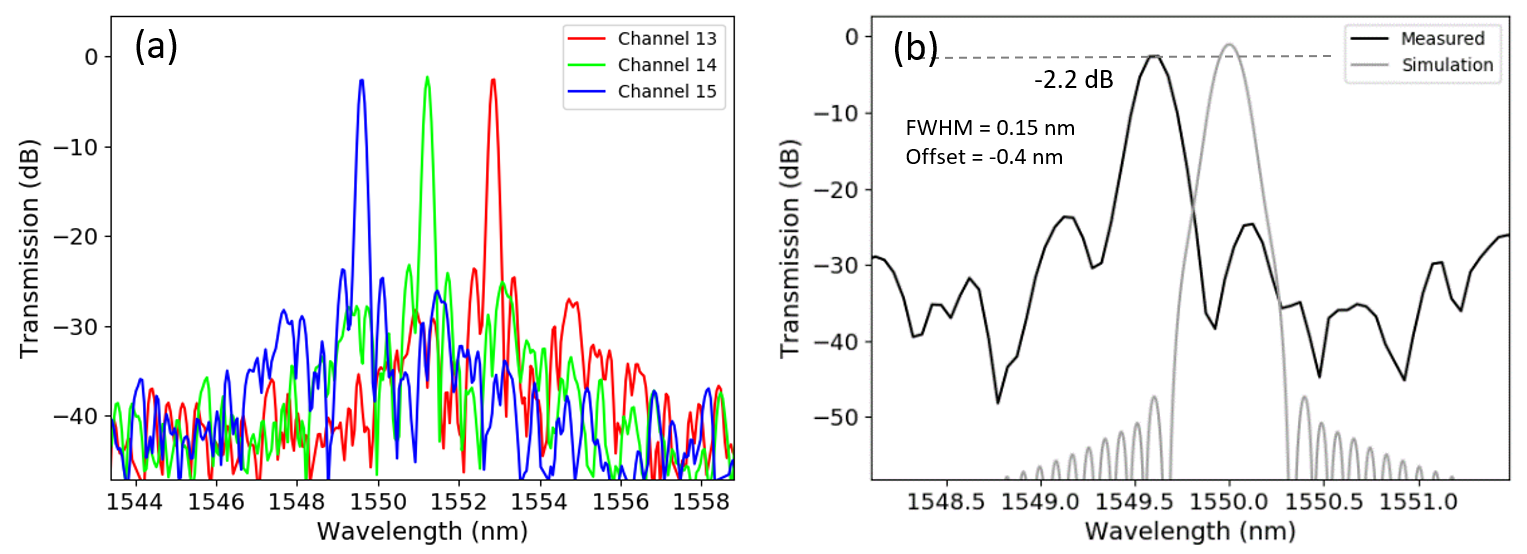}
	\caption{(a) Type E transmission spectrum of output channels 13-15. (b) Main transmission peak in comparison with simulated transmission curve.}
	\label{fig:Meas_TypeE}
\end{figure}

The transmission measurements on types D and E exhibit similar performance with resolving powers of 11,000 and 10,333, respectively. Measurement results for types D and E are shown in Figures \ref{fig:Meas_TypeD} and \ref{fig:Meas_TypeE}, respectively.

Five fabricated silica AWG designs have been tested for insertion loss and transmission peak FWHM. The measurement results were compared with simulated transmission spectra. For type A, the results have shown a high insertion loss of $13$  
dB and significant crosstalk level of $-5$ dB, along with a broadening of the channel transmission curve in the largest AWG design of type A. This design is severely impacted by fabrication tolerances, reaching only $25\%$ of its target spectral 
resolving power. The second group, types B and C, show a similar performance, reaching $57\% - 63\%$ of their theoretical resolving power with an insertion loss of $5$ dB. Designs of the third group, types D and E, showed the best performance 
in terms of insertion loss and agreement with target resolution specifications. Insertion loss was lowest for type E at $2.2$ dB and slightly higher at $2.7$ dB for type D. 
\begin{table}[!ht]
	\caption{\label{tab:exp_summary} Summarized characterization results of AWG designs A-E.}
	\centering
	\begin{tabular}{c c c c c c} 
	 	\hline
		 Type & IL (dB) & Crosstalk (dB) & Bandwidth (nm) & est. $R$ & $\uplambda_0$ offset (nm)\\ [0.5ex] 
		 \hline
		 A & -13 & -5 & 0.1 & 15,500 & -0.19 \\ 
		 B & -5 & -10 & 0.082 & 18,900 & -0.33 \\
		 C & -5 & -10 &  0.09 & 17,222 & 0.11 \\
		 D & -2.7 & -21.3 & 0.14 & 11,000 & 0.3\\
		 E & -2.2 & -22.8 & 0.15 & 10,333 & -0.4\\ [1ex] 
		 \hline
	\end{tabular}
\end{table}
Transmission peak bandwidth measurements yield $0.14$ nm and $0.15$ nm for type D and E, respectively, in fair agreement with the theoretical value of $0.13$ nm. In terms of spectral resolving power, types B and C outperform all remaining 
designs. The measured wavelength offsets of the output channels range from $-0.4$ nm to $0.3$ nm, indicating a global variability of the waveguide effective index on the order of $4\times  10^{-4}$. Table \ref{tab:exp_summary} summarizes the experimental evaluation of the five AWG designs. 
It should be noted at this point that the resolving power estimates obtained from channel power transmission measurements are lower than the actual maximum resolving power of the AWGs when the spectrum is imaged directly. This is owed to the fact that power transmission 
curves are convolutions of the focused beam with the receiver waveguide mode. Deconvolution of the transmission spectrum reveals the focused image and yields a higher resolving power.

\subsection{Dicing of the output facet}
\begin{figure}[!ht]
	\centering
	\includegraphics[width=70mm]{ 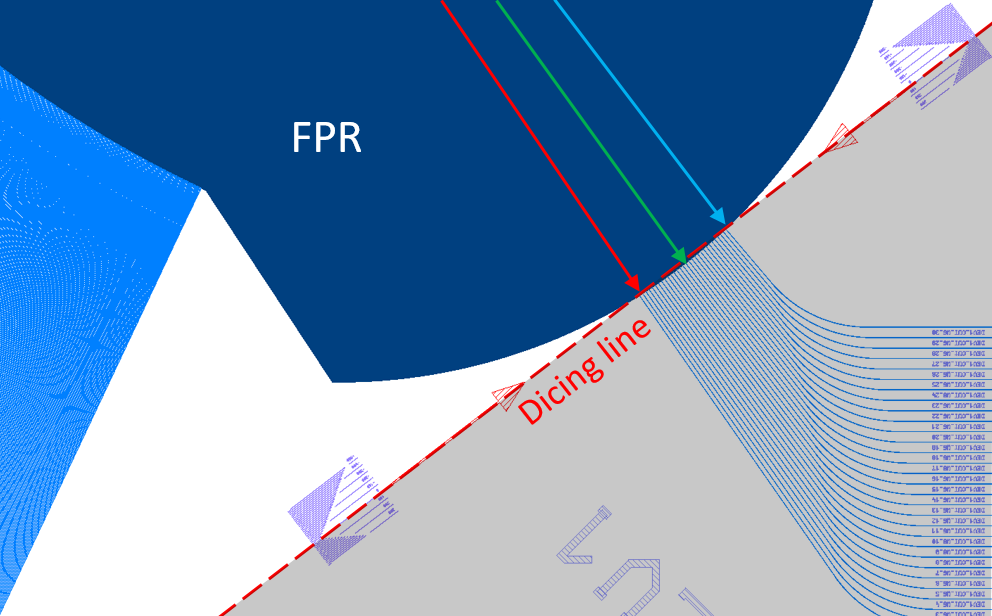}
	\caption{Dicing scheme of the output facet (shown for type E AWG design). Blue, green and red arrows indicate the focused beam locations for shorter and longer wavelengths.}
	\label{fig:AWG_dicing}
\end{figure}

While an array of densely spaced output waveguides can be used for the sampling of the spectrum, the spectral resolution of such a configuration is limited by the integration density of the output waveguides. 
Applications in high-resolution spectroscopy benefit from direct imaging of the spectrum on a flat, polished end facet of the output FPR. Devices of type C and E were selected for further processing due to the high spectral resolution of type C 
and the low insertion loss of type E. Device B, while having a higher resolving power than device C, was retained for further study. The procedure consists of cutting the output facet along a straight line intersecting the FPR just below 
the output waveguides, followed by polishing of the facet to optical quality, schematically shown in Figure \ref{fig:AWG_dicing}. After dicing and polishing, spectral images can be acquired as near-field images of the AWG output facet. 
The numerical aperture of the beams emerging from the AWG is approximately $0.22$, therefore a microscope objective with $NA\geq 0.22$ must be used to avoid vignetting of the beam. Due to the Rowland-geometry of the AWG devices, 
the focus of the output beam is located on a Rowland-circle which intersects the flat polished facet at the edges of the spectral image region. Defocus aberration at the output facet results in spectral resolution imbalance between the edges and 
the centre of the image plane, whereby the central region defines the minimum spectral resolution of the AWG.

In order to fully remove the output waveguides, the output facets of the AWGs were diced and polished $38\, \upmu$m and $47\, \upmu$m beyond the Rowland circle for type C and type E AWGs, respectively, resulting in $1390\,\upmu$m and 
$1625\,\upmu$m wide flat facets. The dicing and polishing procedure was performed at Enablence USA Components. The imaging properties of the processed AWG devices were studied using an imaging system consisting of a Xenix InGaAs NIR 
camera and a microscope objective lens arranged along a single optical axis. Near field images of the AWG output facets were captured in a wavelength range equal to the $FSR$ of each AWG, tuning the TLS wavelength in discrete steps of 
$0.5$ nm. Due to the restricted field of view (FOV) of the microscope objective, only 1/6 of the output facet could be covered by a single exposure, for which reason the camera and objective were moved parallel to the AWG facet to capture 
the region of interest. Optimal focus was achieved by minimizing the spot size in the vertical direction perpendicular to the plane of the FPR. 
\begin{figure}[!ht]
	\centering
	\includegraphics[width=133mm]{ 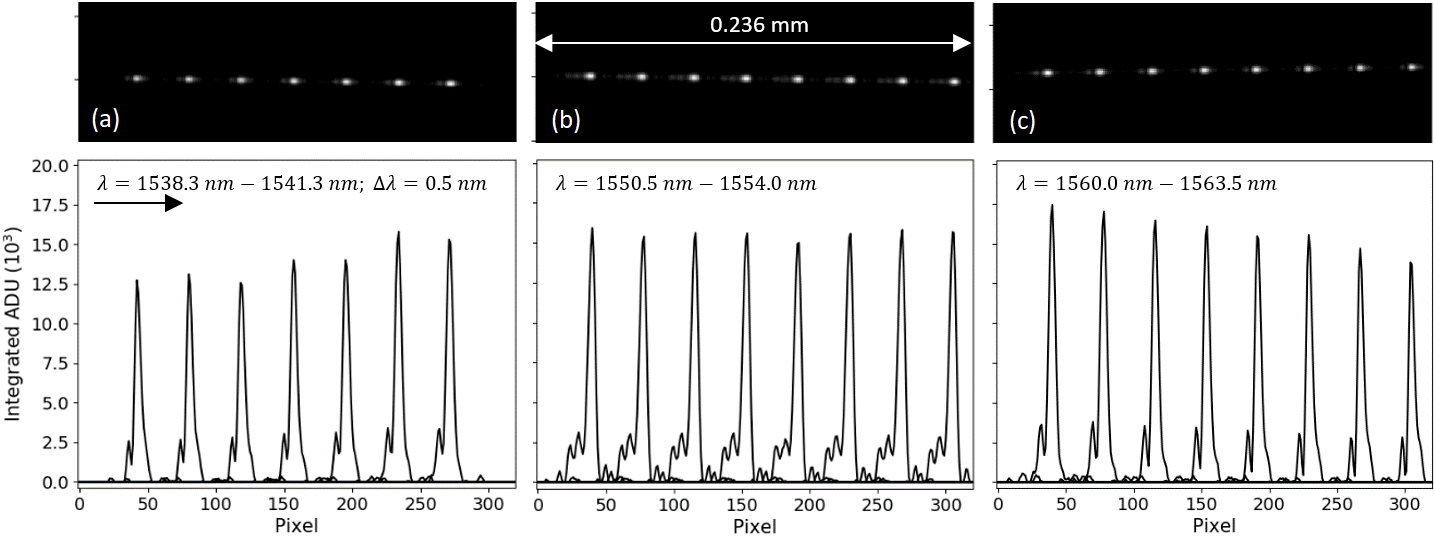}
	\caption{Near field images of the diced output facet of AWG type C. Images taken at the edges (a, c) and centre of the facet (b). The arrow on the left side indicates the direction of increasing wavelength.}
	\label{fig:AWG_typeC_Img}
\end{figure}

The results of the diced facet imaging for AWG type C are shown in Figure \ref{fig:AWG_typeC_Img}. The figure shows three selected segments of the AWG facet: short-wavelength and long-wavelength edges and the central region. Each segment 
shows a superposition of near field images taken in $0.5$ nm wavelength increments. The images on the top show the AWG facets as seen by the camera. Vertical integration of columns was used to process the 2D camera images into 1D vertical slices, 
shown in the plots below the images. The intensity distributions show phase-error induced sidelobes reaching $20\%$ of the main peak maximum. The averaged FWHM of the peaks at the centre of the AWG facet was determined as $4.42\,\upmu$m 
and the wavelength-dependent lateral peak shift was determined as $56.3\,\upmu\text{m/nm}$. A spectral resolving power of $R=19,600$ in the central region of the AWG facet was estimated from the measurement. Similarly, the average peak FWHM 
was determined as $4.09\,\upmu$m and $4.0\,\upmu$m at the short wavelength and long wavelength ends of the facet, respectively, resulting in a spectral resolving power estimate of $R=21,200$ and $R=21,900$ at the edges of the spectral image. 
The reduction of the spectral resolving power towards the centre of the AWG facet is expected due to defocus aberration of the Rowland grating geometry.

\begin{figure}[!ht]
	\centering
	\includegraphics[width=133mm]{ 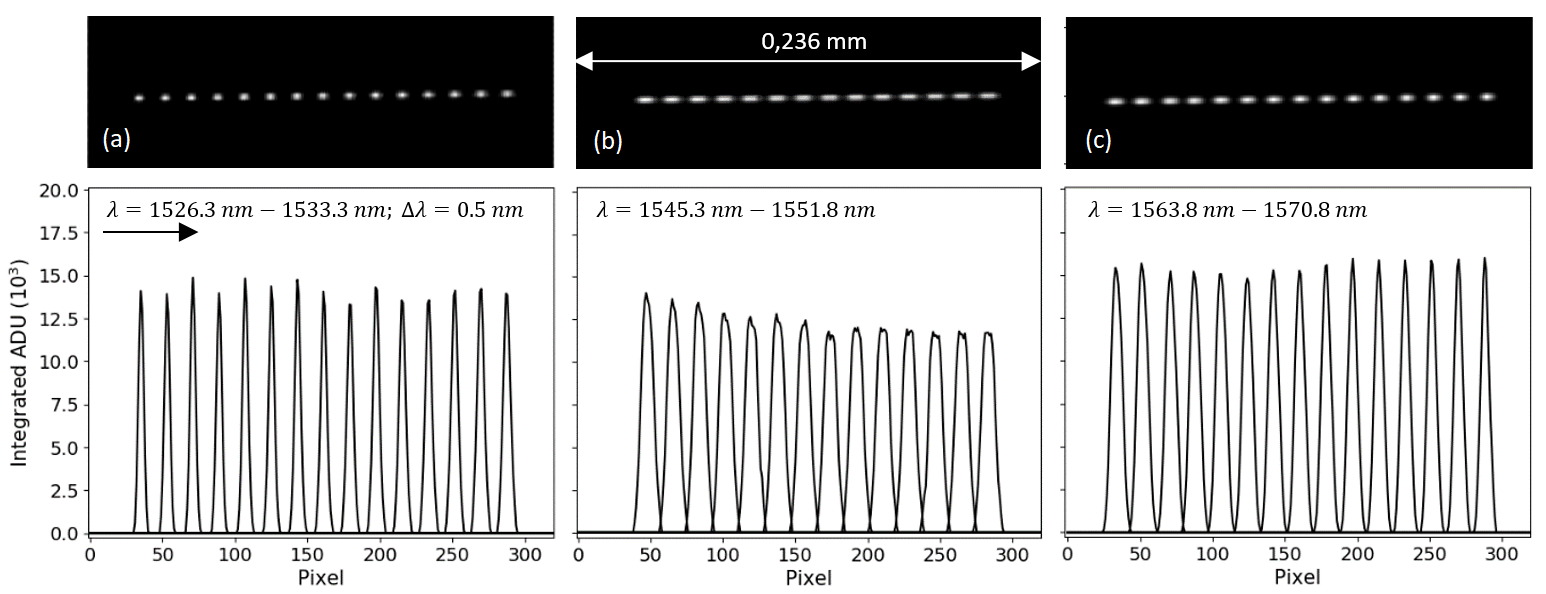}
	\caption{Near field images of the diced output facet of AWG type E. Images taken at the edges (a, c) and centre of the facet (b). The arrow on the left side indicates the direction of increasing wavelength.}
	\label{fig:AWG_typeE_Img}
\end{figure}
\begin{figure}[!ht]
	\centering
	\includegraphics[width=70mm]{ 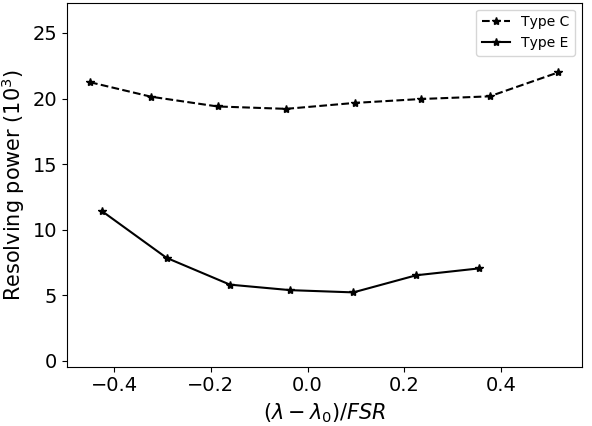}
	\caption{Measured spectral resolving power of type C and type E AWG devices after dicing and polishing of the output facet.}
	\label{fig:Diced_RP}
\end{figure}

The interference image of the diced type E AWG was studied in a similar way using the same experimental setup configuration. Figure \ref{fig:AWG_typeE_Img} shows superimposed near field images in the short-wavelength, central and long-wavelength 
sections of the output facet. While the interference peaks are free of sidelobes, a strong broadening due to defocus aberration can be observed in the central segment of the output facet, indicated by the elongation of the focal spot in horizontal direction. 
Image analysis shows a wavelength-dependent peak shift of $28.5\, \upmu \text{m/nm}$. The peak $FWHM$ varies from $3.9\,\upmu$m at the short wavelength edge between $1526.3$ nm and $1533.3$ nm to $8.47\,\upmu$m in the central region 
between $1551.3\,\text{nm}$ and $1558.8\,\text{nm}$. Accordingly, the spectral resolving power varies between $R=5,215$ in the central region of the image plane and $R=11,320$ at the short-wavelength edge. Figure \ref{fig:Diced_RP} shows 
the variation of measured resolving power for both diced AWG devices on a normalized wavelength axis within the main spectral order. 
Minimum resolving powers of 5,215 (type E) and 19,600 (type C) have been achieved for Rowland-type AWG devices with a diced output image plane. The resolving power non-uniformity was largest in the type E AWG device with a variation of $>50\%$ 
between the edges and the centre of the image plane. 

\begin{figure}[!ht]
	\centering
	\includegraphics[width=130mm]{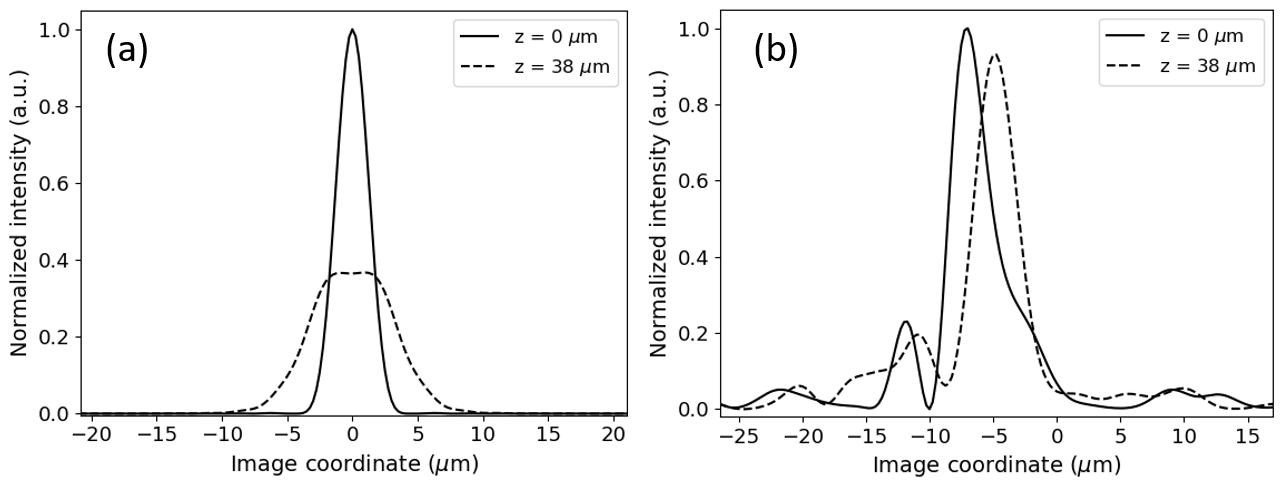}
	\caption{Simulated defocus of the image on the output facet of AWG device C $38\,\upmu$m below the Rowland circle at $1550$ nm. (a) Ideal AWG without path length errors. (b) Fitted model of fabricated device C.}
	\label{fig:Defocus_sim}
\end{figure}
Defocus after dicing was much weaker than expected theoretically for design C, varying only by $10.5\%$ between the edge and centre of the output facet. 
Furthermore, the resolving power measured by direct imaging was higher than the estimate from the transmission measurement in section \ref{sec:trans_measurement}.
This unexpected observation can be explained by distortion of the output beam due to fabrication related optical length errors of the waveguide array. 
We have used a Monte-Carlo method similar to the approach introduced in \cite{Oh:12} to fit a model of the AWG to the measured transmission curve (see Figure \ref{fig:Meas_TypeC}) by random variation of the optical lengths of the array 
waveguides and minimization of the residual sum of squares between the model output and the measured data. After including the fitted data into the AWG simulation, we observed a very similar qualitative behaviour as seen in the experiment. 
In the simulation of the diffraction image at the centre of the polished facet including the fitted path length errors, the FWHM of the intensity peak increased from $3.62\,\upmu$m in the spot of maximum focus at $z=0\,\upmu$m relative to the 
Rowland circle to $3.89\,\upmu$m at a location $38\,\upmu$m below, resulting in a resolving power of 22,400. In contrast, the simulated FWHM increased from $2.91\,\upmu$m at $z=0\,\upmu$m to $7.64\,\upmu$m at $z=38\,\upmu$m in the 
ideal AWG with perfectly tuned path lengths. Simulations of output channel transmission bandwidth of the fitted model almost exactly matched the measured 3-dB bandwidth of 0.09 nm, corresponding to $R=17,130$. 
Figure \ref{fig:Defocus_sim} shows a comparison of defocus behaviour of an ideal AWG of type C with the simulated defocus behaviour of the fabricated AWG.

The near field imaging results show that defocus aberration is a significant performance limiting factor. The problem of defocus aberration is solved in the three-stigmatic-point AWG design \cite{Wang:2001}, which is particularly essential for 
spectroscopy applications. The three-stigmatic-point method modifies the shape of the FPR as well as the lengths of the array waveguides to produce AWG designs with a flat image-plane and minimal aberrations in compliance with the requirements 
of astronomical spectrographs utilizing cross-dispersion. We will present a detailed theoretical and experimental analysis of three-stigmatic-point silica AWG devices in part II of our series.
 
\begin{figure}[!ht]
	\centering
	\includegraphics[width=133mm]{ 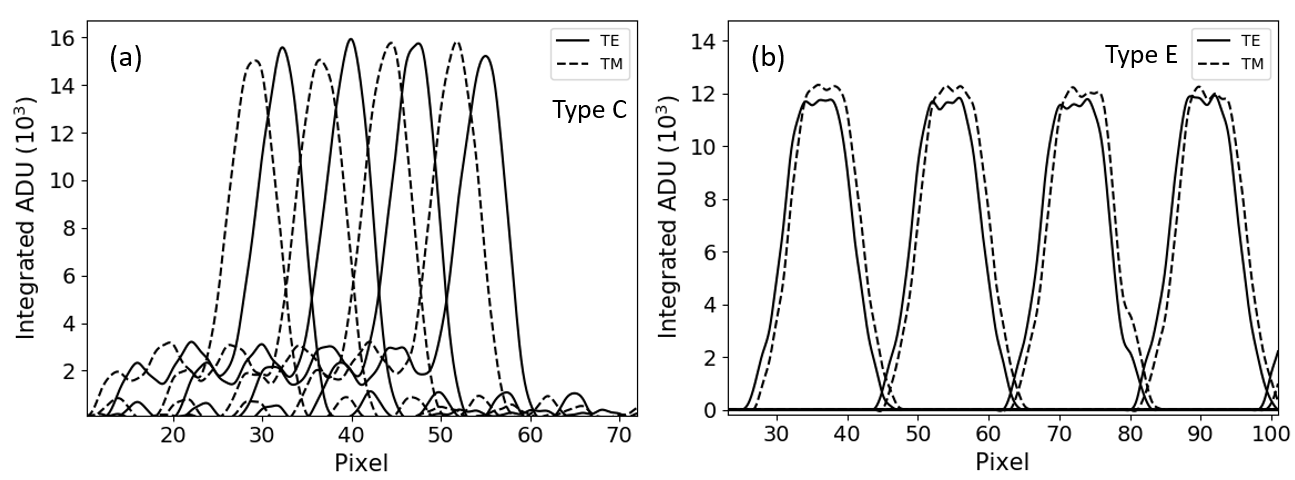}
	\caption{Polarization dependent shift of the image near $1550$ nm. (a) Type C: images taken in wavelength increments of $0.1$ nm. (b) Type E: images taken in wavelength increments of $0.5$ nm.}
	\label{fig:Pol_C_E}
\end{figure}

In the next step, we have studied the polarization sensitivity of the diced AWGs by comparing images of interference peaks taken with TE and TM-polarized input light. Although the waveguide core design aims to minimize geometrical waveguide 
birefringence, the presence of stress-induced birefringence cannot be ruled out. In order to measure the polarization-dependent peak shift in the image plane, we took images at several wavelengths near $\lambda_0 = 1550$ nm with purely 
TE-polarized light and repeated the procedure for the same set of wavelengths with purely TM-polarized light. The superimposed measurement results for type C and type E AWG are shown in Figure \ref{fig:Pol_C_E}.
Figure \ref{fig:Pol_C_E}(a) shows the results for type C AWG. The images were taken in wavelength steps of $0.1$ nm. A strong polarization-dependent shift of the image can be observed. For the type C AWG, we obtain $PD\lambda = 0.04$ nm, 
indicating a waveguide array birefringence $\Delta n_{\text{eff}} = n_{\text{eff,TE}}-n_{\text{eff,TM}}= 3.77\times 10^{-5}$. Similarly, we obtain $PD\lambda=0.026$ nm for the type E AWG, which corresponds to a waveguide birefringence 
$\Delta n_{\text{eff}} = 2.45\times 10^{-5}$. The $PD\lambda$ measured in our experiment is comparable to the value of $0.05$ nm achieved using polarization compensation by core width adjustment in $2.5\%$ silica waveguides \cite{Maru:07}. 
However, a birefringence of the order $10^{-5}$ is still significant considering the high spectral resolution targeted by the AWG designs. Improvement of the AWG designs with regard to polarization sensitivity is subject of future development work. 
Possible methods of birefringence cancellation in AWGs are described in \cite{Inoue:1997, Inoue:2001, Lang:2007}. Birefringence cancellation methods based on an angled star coupler geometry or waveguide core width variation are of special interest, 
as they avoid the post processing and losses introduced by the polyamide half-wave plate method. However, these methods require precise a-priori knowledge of the waveguide birefringence in the fabricated device.

\section{Conclusion}
In the first part of a series of papers, we have presented an experimental study on the practical feasibility of high-resolution ($R=\lambda/\Delta\lambda=15,000...60,000$) arrayed waveguide gratings for use in near-infrared spectroscopy. 
In this study, we have used numerical modeling to develop a set of custom AWG designs, denoted A-E, on a $2\%$ refractive index silica-on-silicon platform for operation in a region of the astronomical H-band covering $1500$ nm - $1700$ nm. 
High-spectral-order AWGs with a narrow free spectral range of $16$ nm as well as low-order AWGs covering up to $48$ nm were designed and characterized.

Output channel transmission spectrum measurements showed increasing degradation of AWG performance with increasing number of waveguides and average waveguide array length due to fabrication-related phase errors. The designs 
with foot-prints between $4.8\,\text{cm}^2$ and $16.38\,\text{cm}^2$ achieved spectral resolving powers between 10,333 and 18,900, estimated from the 3-dB-bandwidth of the output channel transmission, whereby the highest resolution 
was observed in the smallest design. Increasing the spectral resolution significantly beyond $\sim 20,000$ by increasing the focal length and number of waveguides proved to be impractical, as the theoretical benefit was surpassed by degradation 
due to phase errors. Fabricated AWG devices of type C and E were selected for further processing into spectroscopic AWGs by removal of the output waveguides and polishing of the chip facet to optical quality. The interference maximum on the 
polished output facet was directly imaged with a high-NA microscope objective and a NIR camera, using polarization control for TE and TM measurements. Spectral resolving powers of $R \geq 19,600$ (type C) and $R \geq 5,215$ (type E) were 
measured by analysis of the camera images. As expected from AWGs of the Rowland-type, the resolving power was non-uniform across the image plane due to defocus aberration on the flat polished facet and varied by $10.5\%$ in device C 
and $54.4\%$ in device E. The anomalously small defocus in device C was shown to be caused  by phase errors in a numerical simulation of the fabricated device.

Measurements of polarization sensitivity showed a significant PD$\lambda$ of $0.04$ nm for type C and $0.026$ nm for type E. Taking into account the device geometry, a birefringence of $\Delta n_{\text{eff}} = 2.45\times 10^{-5}$ was estimated.

In summary, we have demonstrated the feasibility of custom-developed spectroscopic AWGs achieving spectral resolving powers of up to $R=20,000$ without additional post processing, such as phase error correction by UV trimming \cite{Zauner:98}.
However, we have observed rapid degradation of performance beyond this limit, as phase error induced cross-talk rendered devices with a target resolving power $R\geq 30,000$ practically dysfunctional.
We conclude from this result the necessity of phase error compensation in high-resolution spectroscopic AWGs. Interferometric methods of phase error measurement in AWGs are described in the literature \cite{Takada:00, Takada:06}. Material 
systems other than silica on silicon, such as InP, potentially allow for dynamic phase error correction \cite{Jiang:08}. The presented results show the performance-limiting effect of defocus aberration in Rowland-type AWGs. Therefore, we propose 
implementations of low-aberration three-stigmatic-point geometries in all future generations of spectroscopic AWGs for astronomy. The third major issue is polarization sensitivity of the waveguide array, which can be addressed by modification of the 
waveguide core geometry. These three outlined challenges are essential for future high-performance AWG-based astronomical instruments. The Potsdam Arrayed Waveguide Spectrograph (PAWS), a full cross-dispersion AWG spectrograph demonstrator 
with Teledyne H2RG detector is currently being developed in Potsdam to test the first generation of AWGs as well as future devices on sky \cite{Hernandez:20}. The system will encompass a cryostat in order to minimize thermal radiation in the wavelength 
range of operation as well as the dark current of the image detector, which is essential for long-exposure observations under photon-starved conditions.

\begin{backmatter}
\bmsection{Funding}
This work is supported by the BMBF project “Meta-ZIK Astrooptics” (grant No. 03Z22A511).

\bmsection{Acknowledgments}
This work is supported by the BMBF project “Meta-ZIK Astrooptics” (grant No. 03Z22A511).\\
We gratefully acknowledge the assistance of Dele Zhu, Vadim Makan,  Dr. Julia Fiebrandt and Dr. Ziyang Zhang. 
We especially thank Prof. Dr. Martin Schell, Fraunhofer HHI, Berlin, for sharing general thoughts and providing helpful advice.

\bmsection{Disclosures}
The authors declare no conflicts of interest.

\bmsection{Data availability} Data underlying the results presented in this paper are not publicly available at this time but may be obtained from the authors upon reasonable request.

\end{backmatter}

\bibliography{references}

\end{document}